\documentclass[preprint,showpacs,preprintnumbers,amsmath,amssymb]{revtex4}
 \usepackage{graphicx}
 \usepackage{dcolumn}
 \usepackage{bm}
 \usepackage{ulem}
\usepackage{tensor}

\begin{document}

\title{Possibilities of direct production of superheavy nuclei with $Z=$112--118 in different evaporation channels}

\author{J. Hong,}
\affiliation{Department of Physics and Institute of Physics and Applied Physics, Yonsei University, Seoul 03722, Korea}

\author{G. G. Adamian,}
\affiliation{Joint Institute for Nuclear Research,
Dubna  141980, Russia}

\author{N. V. Antonenko,}
\affiliation{Joint Institute for Nuclear Research,
Dubna  141980, Russia\\
Tomsk Polytechnic University, 634050 Tomsk, Russia}

\author{P. Jachimowicz,}
\affiliation{Institute of Physics,
University of Zielona G\'{o}ra, Szafrana 4a, 65516 Zielona
G\'{o}ra, Poland}

\author{M. Kowal}
\email{michal.kowal@ncbj.gov.pl}
\affiliation{National Centre for Nuclear Research, Pasteura 7, 02-093 Warsaw, Poland}

\date{\today}

\begin{abstract}
The production cross sections of heaviest isotopes of superheavy nuclei with charge numbers
112--118 are predicted in the $xn$--, $pxn$--,  and $\alpha xn$--evaporation channels
of the $^{48}$Ca-induced complete fusion reactions for future experiments.
The estimates of synthesis capabilities are based on a uniform and consistent set of input nuclear data.
Nuclear masses, deformations, shell corrections, fission barriers and decay energies
are calculated within the macroscopic-microscopic approach
for even-even, odd-Z and odd-N nuclei. For odd systems the blocking procedure is used.
To find, the ground states via minimisation and saddle points using Immersion Water flow technique,
multidimensional deformation spaces, containing non-axiallity are used.
As shown, current calculations based on a new set of mass and barriers,
agree  very well with experimentally known cross-sections, especially in the $3n$--evaporation channel.
The dependencies of these predictions on the mass/fission barriers tables and fusion models are discussed.
A way is shown to produce directly unknown superheavy isotopes in the $1n$-- or $2n$--evaporation channels.
The synthesis of new superheavy isotopes unattainable in reactions with emission of neutrons is proposed in the promising channels
with emission of protons ($\sigma_{pxn} \simeq 10-200$ fb) and alphas ($\sigma_{\alpha xn} \simeq 5-500$ fb).

\end{abstract}

\maketitle

\section{Introduction}

The production and spectroscopic study of superheavy nuclei (SHN) is currently one of
the important topics in nuclear experiment and theory.
Due to the extremely short lifetimes of SHN and the exceptionally low probabilities
of their production the final cross-sections are extremely small.
Although determine them is a huge challenge for today theory,
only having reached this stage it will be be possible to make reliable predictions of probabilities for synthesis
even heavier, still non-existent SHN.
The $^{48}$Ca-induced
complete fusion reactions have been  successfully  used to synthesize
SHN with the charge numbers $Z$=112--118 in
the neutron evaporation   channels ($xn-$evaporation channels)
\cite{Og1,Og2,Og1n,SH1,SH2,SH3,SH4,SH5,SH6,SH7,SH8}
and to approach to "the island of stability" of SHN predicted
at $Z$=114--126 and neutron numbers $N$=172--184 by the nuclear
shell models \cite{Sobi,Sobimasstable,MF,HofMun}.
The most of these SHN have been obtained in the  $3n-$ and $4n-$evaporation channels.
Only in the reactions $^{48}$Ca+$^{242}$Pu, $^{48}$Ca+$^{243}$Am, and $^{48}$Ca + $^{245}$Cm
the evaporation residues have been detected in the $2n$-evaporation channel. The nuclei  $^{285,287}$Fl and  $^{292}$Ts
have been also produced in the $5n$-evaporation  channel.
On the agenda is to expand the region of SHN in the
direction of the magic neutron number $N=184$,  the center of
the predicted "island of stability". For this purpose, we should study both new experimental possibilities and
possible reaction channels.
New isotopes of heaviest nuclei with $Z$=112--117 can   be synthesized in the $^{48}$Ca-induced  actinide-based complete
fusion-evaporation reactions with the emission of charged particle (proton "$p$" or alpha-particle "$\alpha$") and neutron(s) from the compound nucleus (CN).
Note that the possibility of the production of new heaviest isotopes of superheavy nuclei with charge numbers 113, 115, and 117
in the proton evaporation channels with rather high efficiency was suggested for the first time in Ref. \cite{OgCharge}.
This extremely intriguing suggestion was tested in Refs. \cite{model2} and \cite{SWCK}.

One can also observe new isotopes in the $1n-$ and $2n-$evaporation
channels of the$^{48}$Ca-induced  actinide-based complete fusion reactions.
Using the predictions of superheavy nuclei properties of Ref. \cite{moller}, we have recently studied
these possibilities in Refs. \cite{model2,model3}. We have revealed how rapidly  the evaporation
residue  cross section  decreases with decreasing beam energy in the  sub-barrier region.

An interesting question is how the estimation of production cross sections change if we replace the mass table containing
the predictions of SHN properties. Taking other mass table, we should incorporate it in all
steps of the calculation of the evaporation residue cross sections. As known, the evaporation residue cross sections
depend on the capture cross section, fusion probability (formation of the CN), and survival probability
(the survival with respect to fission). The last one seems to be the most sensitive to the SHN properties.
However, the fusion probability also crucially depends on the change of the mass table because it
affects the potential energy surface driving two colliding nuclei to the CN. The capture
cross section depends on the deformations predicted for the colliding nuclei.
So, in the present article, employing the mass table of Ref.~\cite{MKowal, Jach2014, Jach2017} based on the microscopic-macroscopic method,
 we will predict the  excitation functions in   the $xn-$, $pxn-$,  and $\alpha xn-$evaporation channels
of the $^{48}$Ca-induced  complete  fusion   reactions and, correspondingly, the maximum cross
sections at the optimal energies of these channels.

\section{Model}

For the excited SHN, the emission of
charged particles is suppressed by the high Coulomb barrier and
 competes with the neutron evaporation and fission.
The evaporation residue cross section can be written in factorized form
\cite{model2,model3,AA,AAIS,AA2,AA9,Avaz,Acta,Bao,Wang1,Wang2,Wang3,dns,charge,lecture,model,paper1}:
\begin{equation}
\sigma_{s}(E_{\rm c.m.}) =\sum_{J=0}
\sigma_{cap}(E_{\rm c.m.},J)P_{CN}(E_{\rm c.m.},J)W_{s}(E_{\rm c.m.},J).
\label{ER_eq}
\end{equation}
The evaporation residue cross section in the evaporation channel $s$ depends on the partial
capture cross section $\sigma_{cap}$
for the transition of the colliding nuclei over the entrance
(Coulomb) barrier, the probability of CN
formation $P_{CN}$ after the capture and the survival probability
$W_{s}$ of the excited CN.
The formation of CN is described within a version of the dinuclear system model \cite{model2,model3,model,paper1}.

 In the first step of
fusion reaction the projectile is captured by the target.
In the calculation of $\sigma_{cap}$ in Eq. (\ref{ER_eq}), the orientation of the
deformed actinide target nuclei is taken into account \cite{model}.
The bombarding energy $E_{\rm c.m.}$ at which the capture for all orientation becomes possible is
defined by the Coulomb barrier at sphere-side orientation. At smaller $E_{\rm c.m.}$ some partial waves fall
under the barrier. The position and height of the Coulomb barrier are mainly affected by the quadrupole
deformation of actinide nucleus. The quadrupole
deformation used  were extracted in Ref. \cite{Raman} from the measured quadrupole moments. So, the
effect of deformations of higher multipolarities is taken partially into consideration in our calculations.
Because the uncertainty in quadrupole deformation affects the Coulomb barrier stronger than the hexadecopole
deformation, we consider only  quadrupole
deformation in our calculations.

In the second step the formed dinuclear system (DNS) evolves to the
CN in the mass asymmetry coordinate $\eta=(A_1-A_2)/(A_1+A_2)$ ($A_1$ and $A_2$ are
the mass  numbers of the DNS nuclei)
\cite{model2,model3,AA,AAIS,AA2,AA9,Avaz,Bao,Wang1,Wang2,Wang3,dns,charge,lecture,model,paper1}.
Because the bombarding energy $E_{\rm c.m.}$
of the projectile is usually higher than the $Q$ value for the
CN formation, the produced CN is excited.

When successful, hot fusion creates a heavy nucleus in a highly excited state that rapidly emits two, three
or four fast neutrons, each removing a few MeV of energy from the system, whereby it “cools
down.” At every stage of this emission the neutrons
compete with the fission processes that lead to nucleus
splitting. This means that the nucleus generated
through hot fusion must be resistant to nuclear fission and explain great importance of the fission barrier $B_{f}$
as the main parameter which protect nucleus against fission. In another words, in the third step of the reaction the CN loses its
excitation energy mainly by the emission of particles and $\gamma-$quanta
\cite{BT,VH,Ignb,wsur,SM,IL,wsur2,rp}. In the de-excitation of CN, the particle emission
competes with the fission which is the most probable process besides fission in normal nuclei.
In this paper, we  describe the production of nuclei in the evaporation
channels with emission of charged particle  (proton or $\alpha$-particle) and neutrons as in Refs.~\cite{model2,model3,paper1}.
The emissions of $\gamma$, deutron, triton, and clusters heavier than alpha-particle are expected to be
negligible to contribute to the total width of the CN decay.
The de-excitation of the CN is treated with the
statistical model using the level densities from the Fermi-gas model.
The neutron separation energies $B_n$, $Q-$values for proton and alpha-particle emissions,  the nuclear mass excesses of SHN, and
the fission barriers for the nuclei considered  are taken from the microscopic-macroscopic model~\cite{MKowal, Jach2014, Jach2017}.
Recently, within this approach (with parameters adjusted
 to heavy nuclei \cite{WSpar}), it was possible to reasonably reproduce data on
 ground state masses;  first,  second and third \cite{Kow,kowskal,IIbarriers,IIIbarriers1,IIIbarriers2,Jach2014,actinides}
 fission barriers in actinides nuclei for which some emipirical/experimental data are available.
\begin{table}
\centering
\caption{
The theoretical barriers $V_i$
and energy thresholds $B_i=V_i-Q_i$ in the evaporation
channels with emission of   proton and alpha-particle ($i=p,\alpha$). The $Q_{p,\alpha}-$values
are calculated with the microscopic-macroscopic models  \cite{moller} and  \cite{MKowal}.
}
\newcolumntype{C}{>{\centering\arraybackslash}p{17.5ex}}
\begin{tabular}{|c|c|c|c|c|c|c|}
\hline
Reaction             & $V_p$  & $V_\alpha$ & $B_p$ \cite{MKowal} &  $B_\alpha$ \cite{MKowal} & $B_p$ \cite{moller} &  $B_\alpha$  \cite{moller} \\
                     &(MeV) &   (MeV)    & (MeV)        &   (MeV)   & (MeV)        &   (MeV)   \\
\hline
$^{48}$Ca+$^{242}$Pu  &  12.6  &  25.1 & 16.7  & 15.0   & 17.1  & 16.6 \\
$^{48}$Ca+$^{244}$Pu  &  12.6  &  25.1 & 16.1  & 15.7   & 17.2  & 16.9 \\
$^{48}$Ca+$^{243}$Am  &  12.7  &  25.3 & 14.5  & 14.1   & 14.1  & 15.7 \\
$^{48}$Ca+$^{245}$Cm  &  12.8  &  25.5 & 15.5  & 14.7   & 15.5  & 14.6 \\
$^{48}$Ca+$^{248}$Cm  &  12.7  &  25.5 & 16.1  & 14.6   & 15.9  & 14.4 \\
$^{48}$Ca+$^{249}$Bk  &  12.8  &  25.6 & 14.0  & 14.1   & 14.2  & 13.9 \\
$^{48}$Ca+$^{249}$Cf  &  12.9  &  25.9 & 15.0  & 13.6   & 14.8  & 13.8 \\
$^{48}$Ca+$^{251}$Cf  &  12.9  &  25.9 & 15.7  & 14.0   & 15.1  & 13.3 \\
\hline
\end{tabular}
\label{barriers}
\end{table}

\begin{figure}
\includegraphics[width=0.45\textwidth,clip]{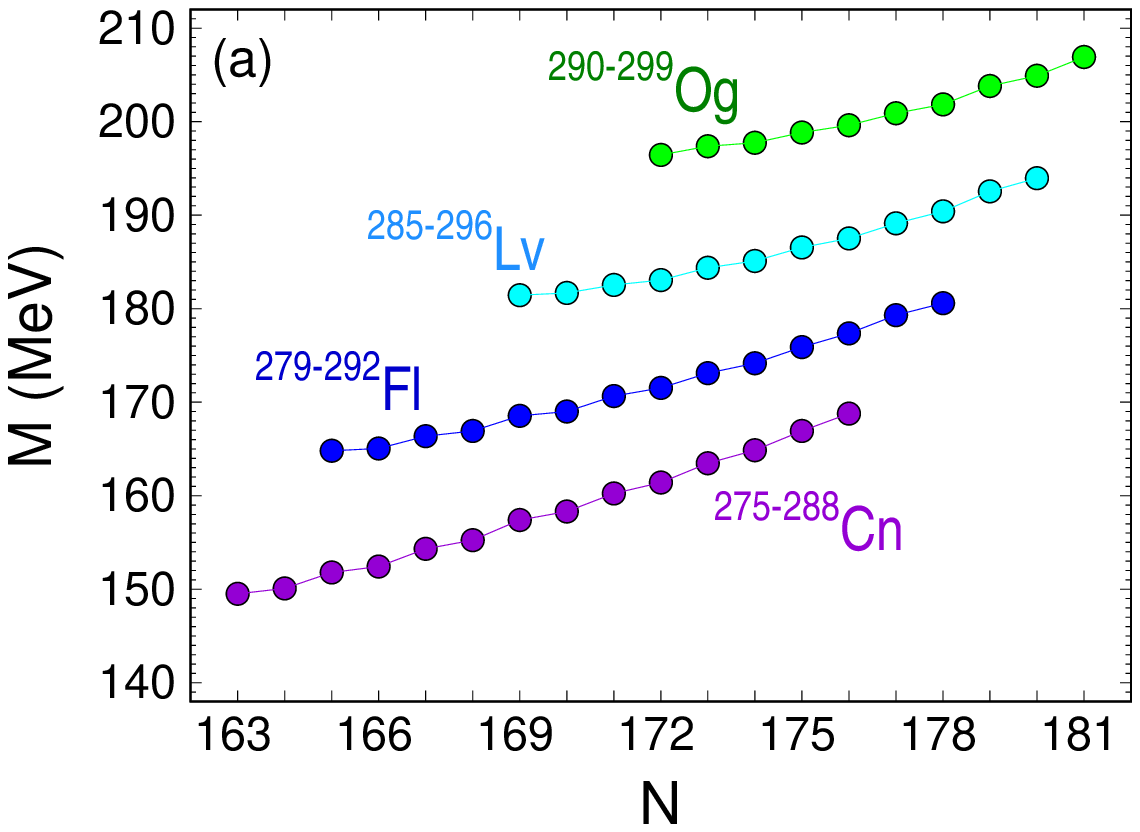}
\includegraphics[width=0.45\textwidth,clip]{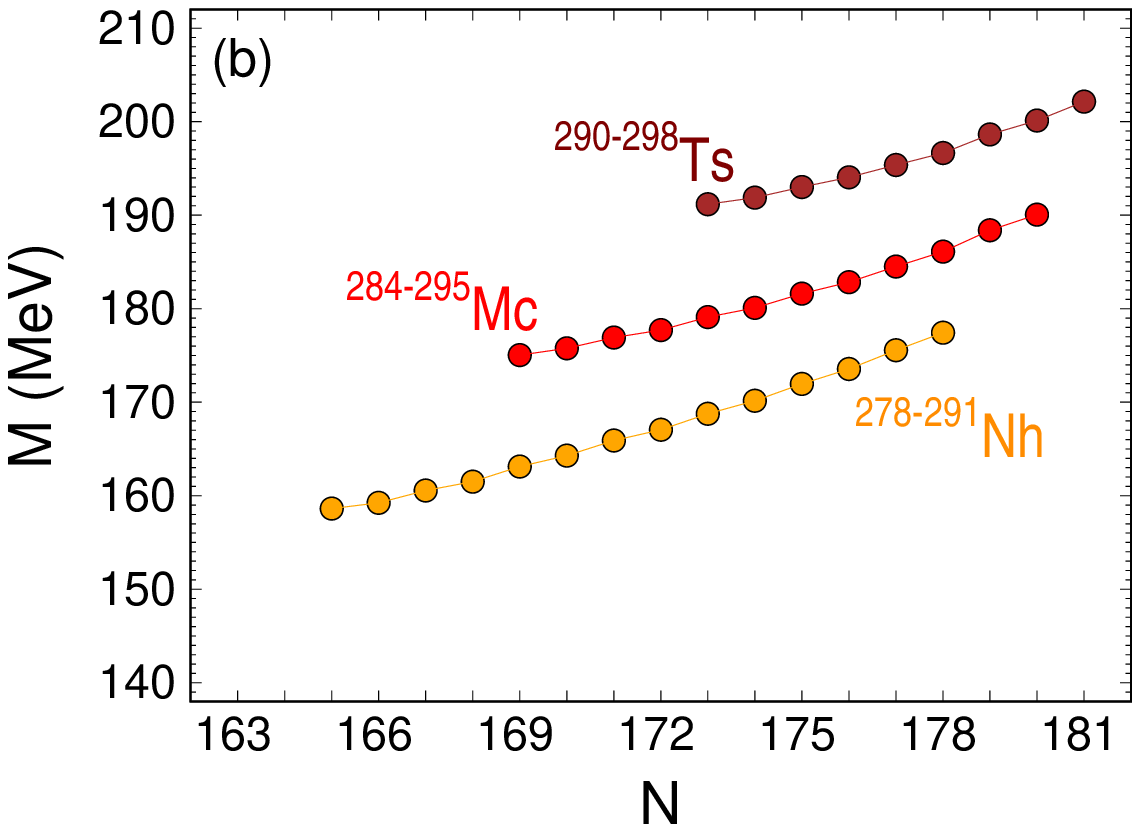}
\includegraphics[width=0.45\textwidth,clip]{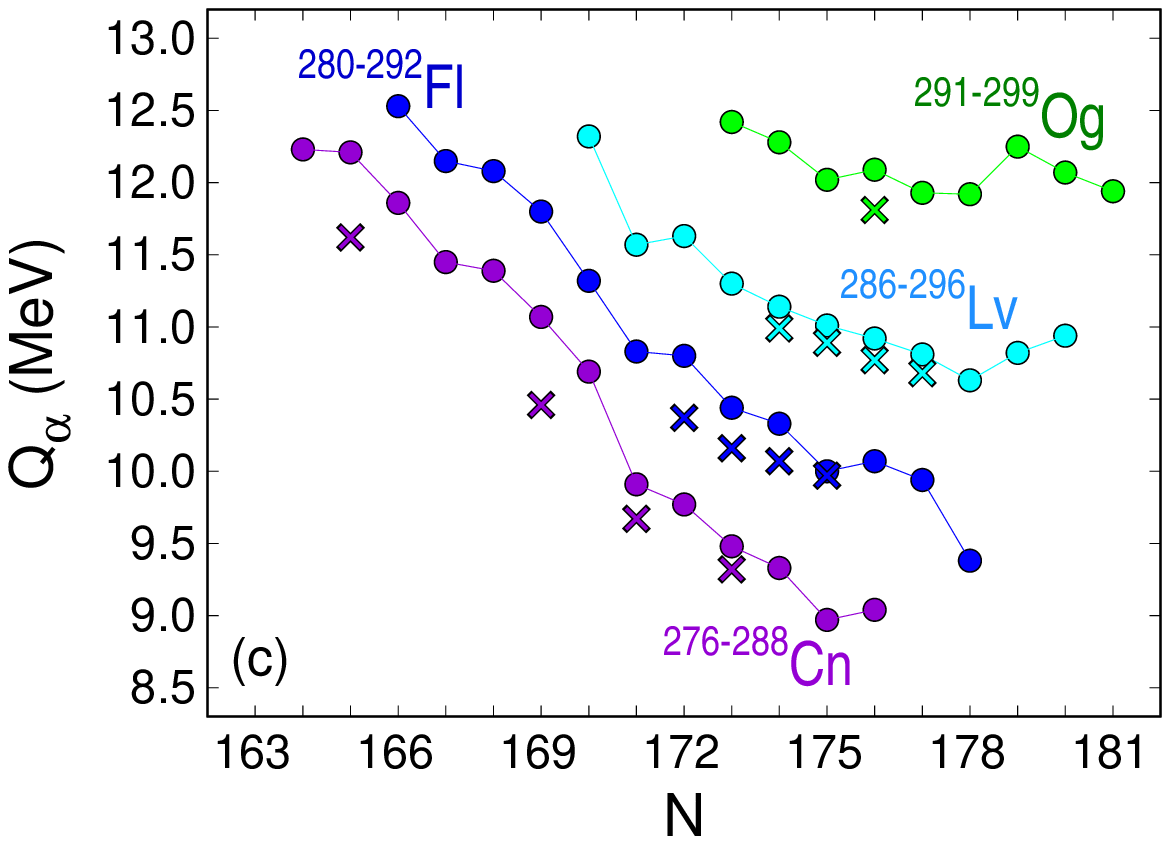}
\includegraphics[width=0.45\textwidth,clip]{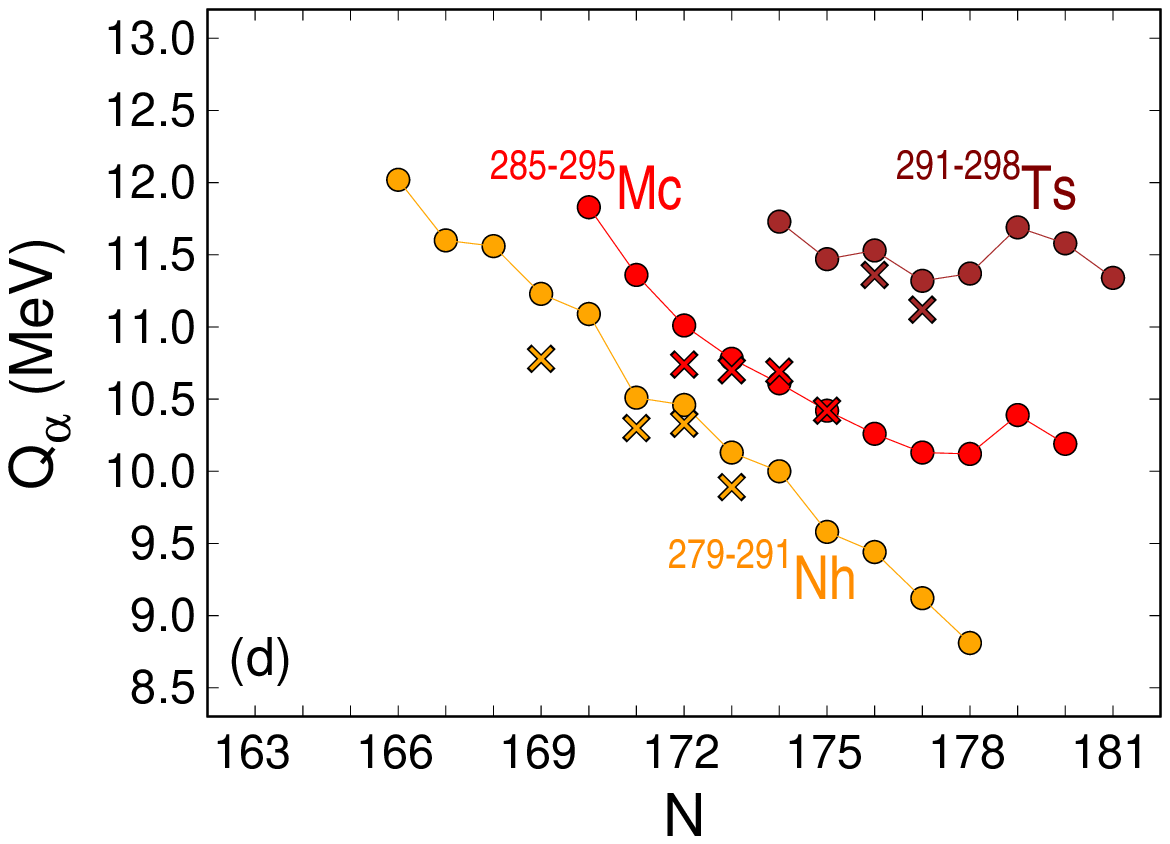}
\caption{
Nuclear masses: $M$ (top panels) and $Q_\alpha$-values (bottom panels) for even-$Z$ (left hand-side panels):
$^{275-288}$Cn, $^{279-292}$Fl,
$^{286-296}$Lv, $^{291-299}$Og
and odd-$Z$ (right hand-side panels): $^{279-291}$Nh, $^{285-295}$Mc, $^{291-298}$Ts.
Experimental data for $Q_\alpha$ are
taken from \cite{Og1n}.
}
\label{masses}
\end{figure}
\begin{figure}
\includegraphics[width=0.45\textwidth,clip]{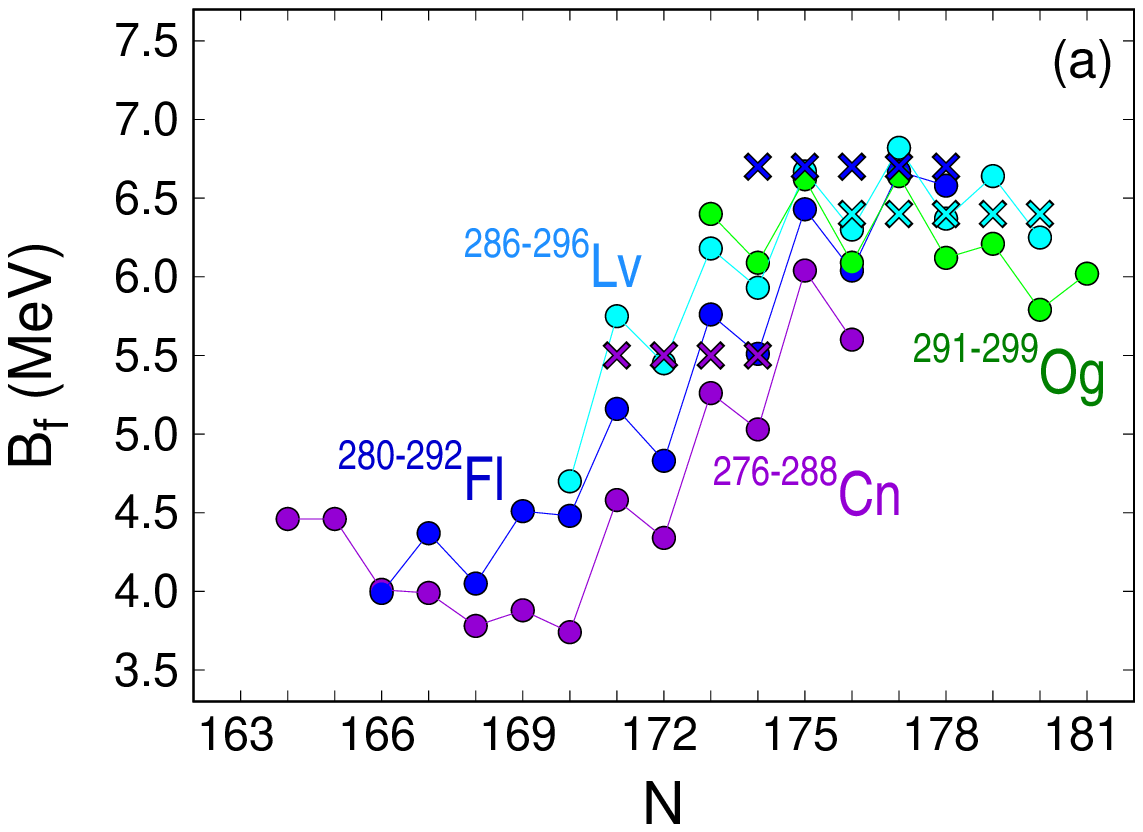}
\includegraphics[width=0.45\textwidth,clip]{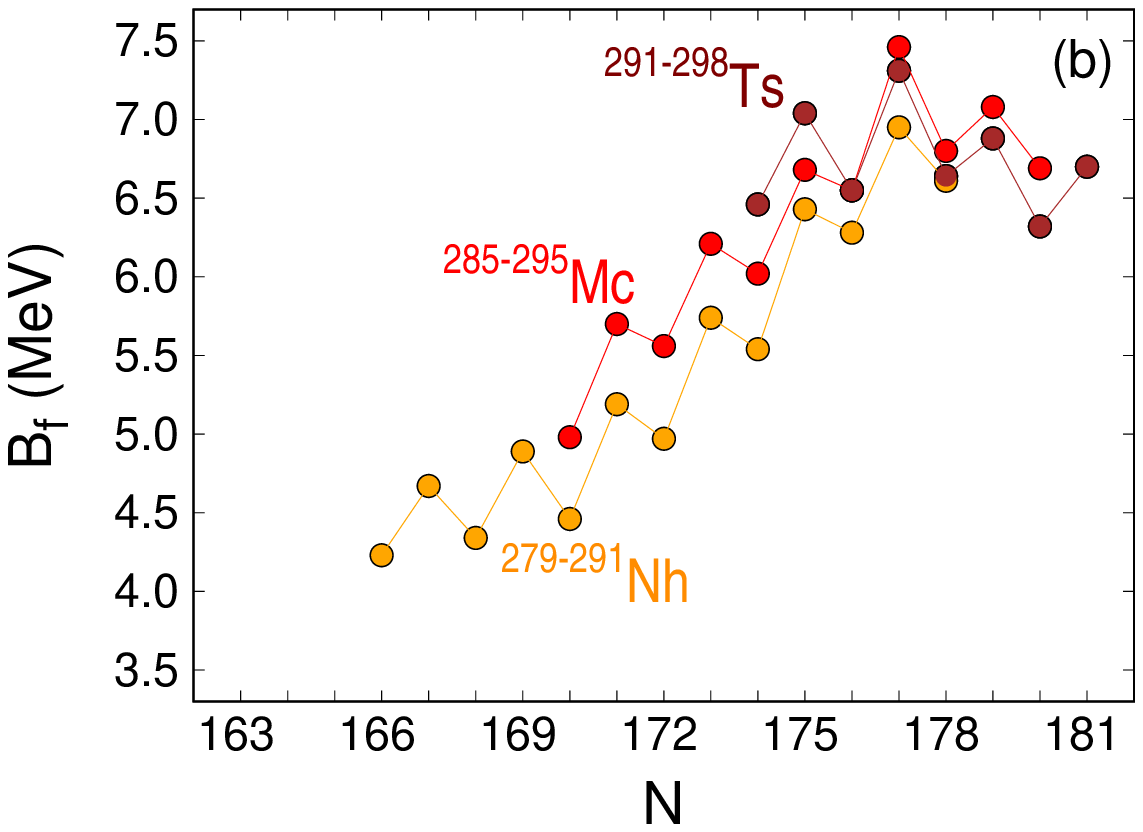}
\includegraphics[width=0.45\textwidth,clip]{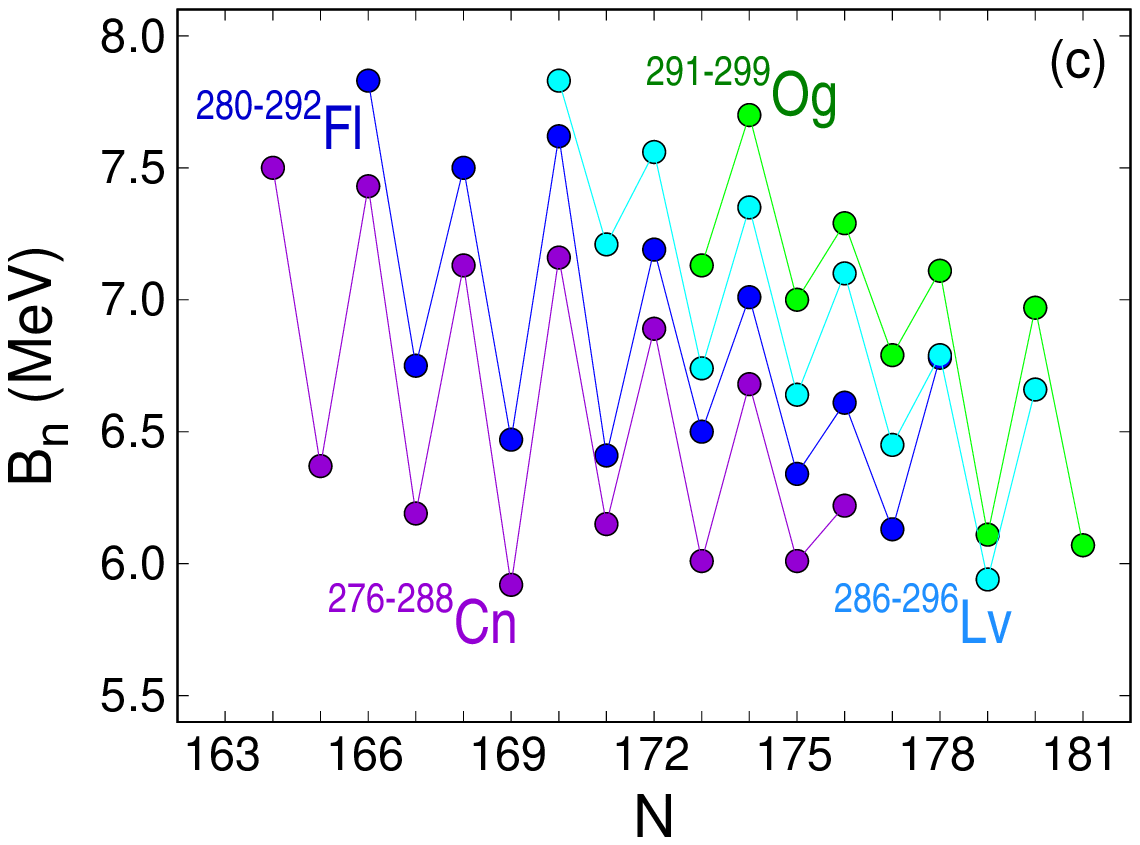}
\includegraphics[width=0.45\textwidth,clip]{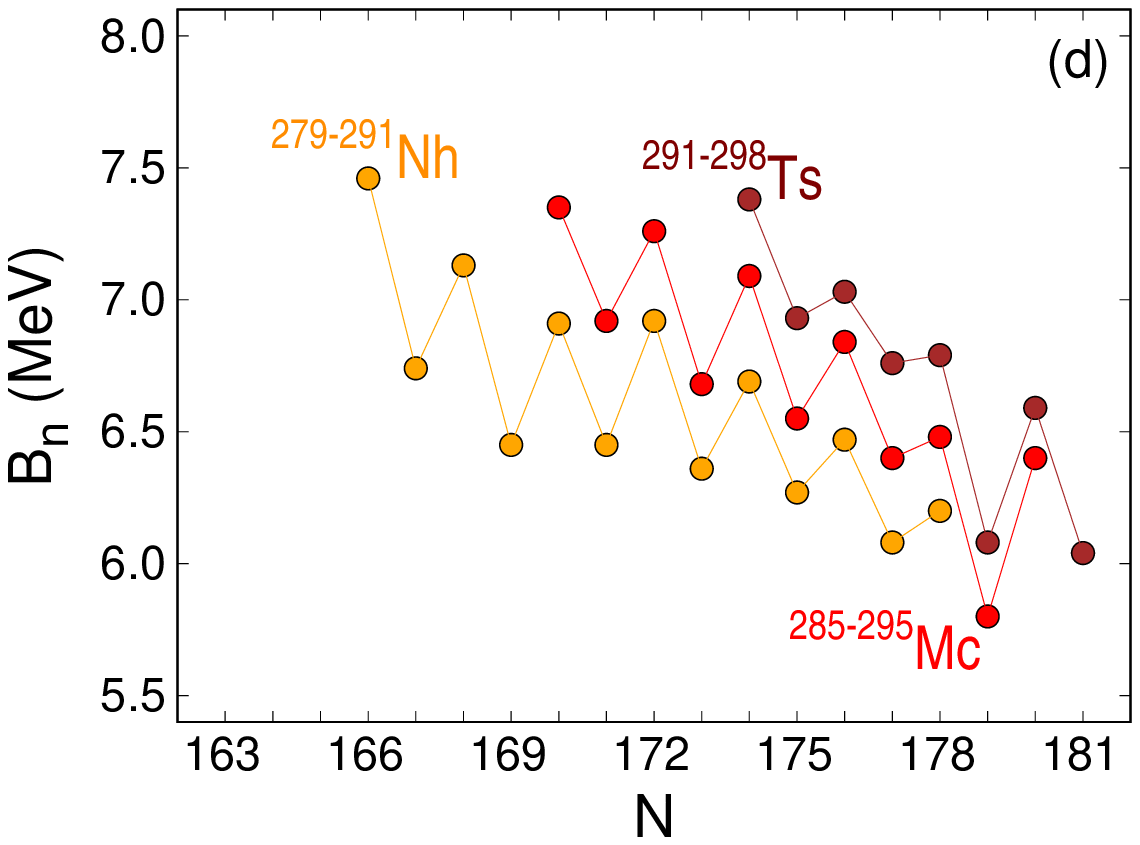}
\includegraphics[width=0.45\textwidth,clip]{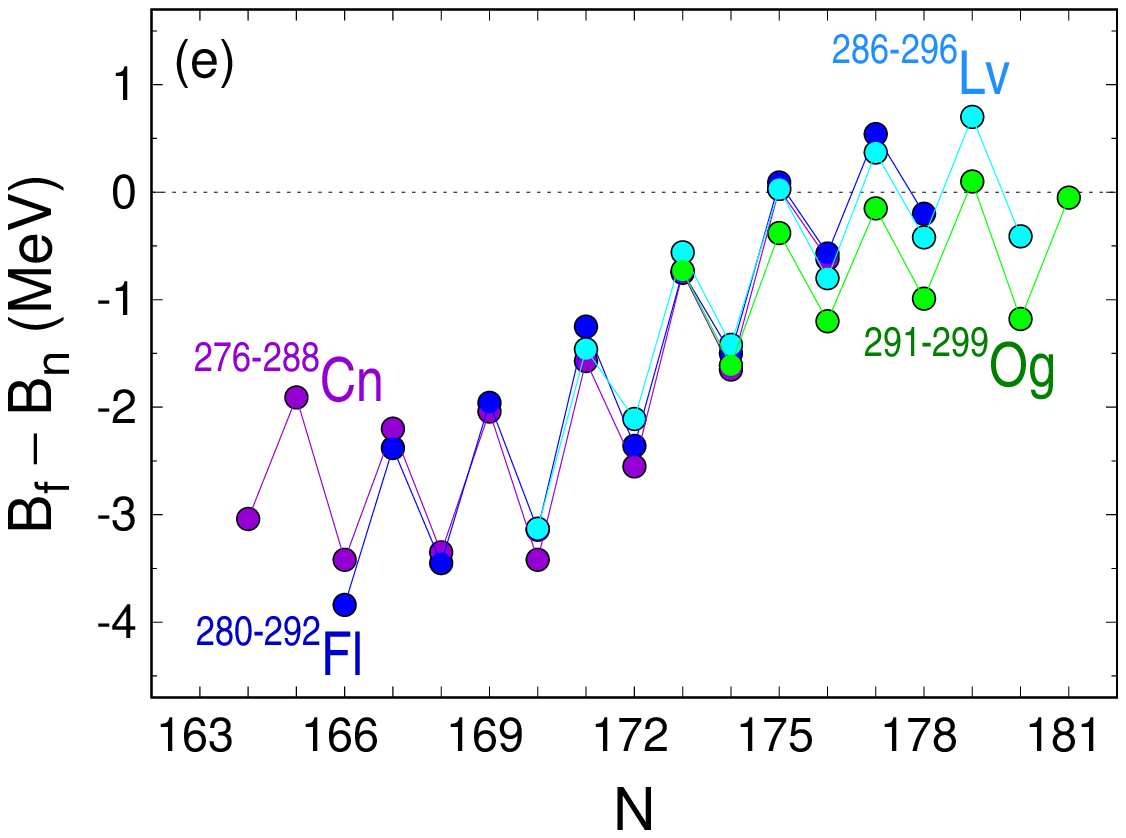}
\includegraphics[width=0.45\textwidth,clip]{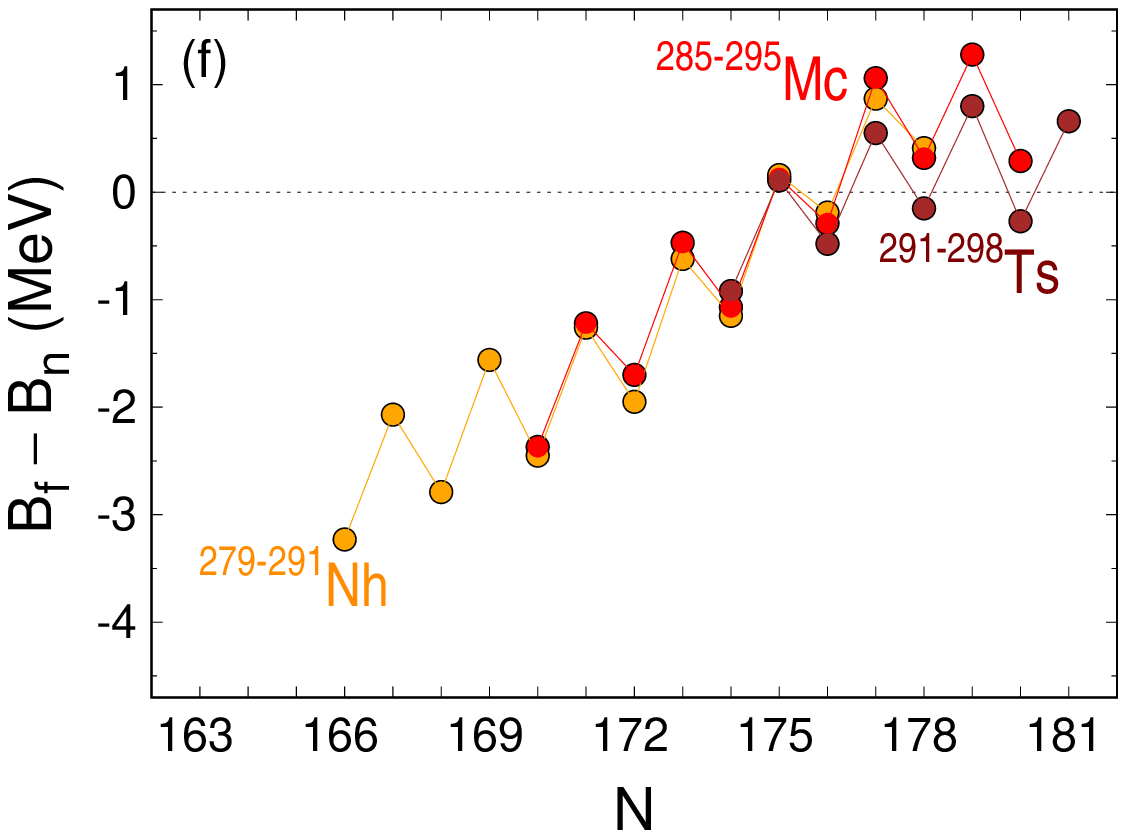}
\caption{
Fission barriers: $B_{f}$ (top panels), neutron separation energies: $B_{n}$ (middle panels) and their differences:
$B_{f}-B_{n}$ (bottom panels) for even-$Z$ (left-hand side panels): $^{275-288}$Cn; $^{279-292}$Fl; $^{286-296}$Lv; $^{291-299}$Og
and odd-$Z$ (right-hand side panels): $^{279-291}$Nh; $^{285-295}$Mc; $^{291-298}$Ts. Experimental data for barriers,
marked by crosses, are taken from \cite{ITKIS}.
}
\label{fisbar}
\end{figure}

\begin{table}[!htbp]
  \caption{Calculated   $Q_\alpha$-values \cite{MKowal} and
  experimental  $Q^{exp.}_\alpha$-values \cite{Og1n} for the indicated SHN.}
  \label{tab:Spect_L}
  \centering
\begin{tabular}{|c|c|c|c|c|c|}
\hline
SHN             & $Q_\alpha$  & $Q^{exp.}_\alpha$ &    SHN             & $Q_\alpha$  & $Q^{exp.}_\alpha$ \\
               &(MeV)         &   (MeV)           &                    &(MeV)        &   (MeV)           \\
\hline
$^{294}$Og  &  12.09  &  11.81 &  $^{288}$Fl  &  10.32  &  10.07 \\
$^{294}$Ts  &  11.25  &  11.12 &  $^{287}$Fl  &  10.44  &  10.16 \\
$^{293}$Ts  &  11.45  &  11.36 &  $^{286}$Fl  &  10.80  &  10.37 \\
$^{293}$Lv  &  10.80  &  10.68 &  $^{286}$Nh  &  9.68  &  9.89 \\
$^{292}$Lv  &  10.92  &  10.77 &  $^{285}$Nh  &  10.27  &  10.33 \\
$^{291}$Lv  &  10.89  &  10.89 &  $^{284}$Nh  &  10.34  &  10.30 \\
$^{290}$Lv  &  11.14  &  10.99 &  $^{282}$Nh  &  10.71  &  10.78 \\
$^{290}$Ms  &  10.11  &  10.42 &  $^{278}$Nh  &  11.56  &  11.85 \\
$^{289}$Ms  &  10.56  &  10.69 &  $^{285}$Cn  &  9.35  &  9.32 \\
$^{288}$Ms  &  10.54  &  10.70 &  $^{283}$Cn  &  9.91  &  9.67 \\
$^{287}$Ms  &  10.61  &  10.74 &  $^{281}$Cn  &  10.78  &  10.46 \\
$^{289}$Fl  &  9.93  &  9.97   &  $^{277}$Cn  &  11.66  &  11.62 \\
\hline

\end{tabular}
\label{barriers}
\end{table}
Within the microscopic-macroscopic method, the energy of deformed nucleus is
 calculated as a sum of two parts: the macroscopic one being a smooth
 function of $Z$, $N$ and deformation, and the fluctuating microscopic one
  that is based on some phenomenological single-particle potential.
 The deformed Woods-Saxon potential model used here
 is defined in terms of the nuclear surface.
Mononuclear shapes can be parameterized via nuclear radius expansion in spherical harmonics ${\rm Y}_{lm}(\vartheta ,\varphi)$.
We admit shapes defined by the following equation of the nuclear surface:
    \begin{equation}
   R(\vartheta ,\varphi)= c(\{\beta\}) R_0
 \{ 1+ \sum _{\lambda=1}^{\infty}\sum _{\mu=-\lambda}^{+\lambda} \beta_{\lambda\mu}{\rm Y}_{\lambda\mu} (\vartheta ,\varphi)\},
   \label{eq:radius}
\end{equation}
  where $c(\{\beta\})$ is the volume-fixing factor and $R_0$ is the radius of a spherical nucleus.

The $n_{p}=450$ lowest proton levels and $n_{n}=550$ lowest
neutron levels from $N_{max}=19$ lowest shell of the deformed oscillator
 are taken into account in the diagonalization procedure.
 We have determined the single--particle spectra for every investigated
nucleus.
The Strutinsky smoothing was performed with the
6-th order polynomial and the smoothing parameter equal to $1.2 \hbar\omega_0$.
For the systems with odd proton or neutron (or both), we use the standard
blocking method. Considered configurations consist of an odd particle
occupying one of the levels close to the Fermi level and the rest of
particles forming paired BCS state on remaining levels.
The ground states were found by minimizing over configurations (blocking
 particles on levels from the 10-th below to 10-th above the Fermi level)
 and deformations. For nuclear ground states it was possible to confine
 analysis to axially-symmetric shapes. More details can be found in Ref.~\cite{Jach2014}.
The simplest extension of the WS model to odd nuclei required three new
  constants which may be interpreted as the mean pairing energies for even-odd, odd-even and odd-odd nuclei
 \cite{Jach2014}.
 They were fixed by a fit to the masses with $Z\geq82$ and $N>126$ via
 minimizing the rms deviation in particular groups of nuclei what is
 rather standard procedure \cite{mol95,mol97}.
 The experimental nuclear masses of heavy nuclei were taken from \cite{Wapstra2003}.
 The obtained rms deviation in masses for 252 nuclei is about
 400 keV with blocking scenario \cite{Jach2014} used here.
 Similar rms error is obtaimed for 204 $Q_{\alpha}$ values. For 88 measured
 $Q_{\alpha}$ values in SHN, the quantities outside the region of the
 fit, we obtained the rms deviation of about 250 keV \cite{Jach2014}.

To estimate the survival probability, the fission barriers from adiabatic scenario, i.e. the smallest possible ones,
are taken \cite{Jach2017}.
The main problem in a search for saddle points is that, since they are neither
 minima nor maxima, one has to know energy on a multidimensional grid of
  deformations (the often used and much simpler method of minimization with
  imposed constraints may produce invalid results)
 \cite{Moller2009,Dubray,IIbarriers,Schunck}.
  To find saddles on a grid we used the Imaginary Water Flow technique.
 This conceptually simple and at the same time very efficient (from a numerical
  point of view) method was widely used and discussed
 \cite{Luc91,Mam98,Hayes00,Moeler04,Moller2009,IIbarriers}.
Based on this and our previous results showing that triaxial saddles are abundant in SHN \cite{Kow}, we consider that quadrupole
triaxial shapes have to be included for the first barriers with which we are dealing with the nuclei considered here.
So, the saddle points are searched in the five dimensional deformation space spanned by:
 $\beta_{20}$, $\beta_{22}$, $\beta_{40}$, $\beta_{60}$, $\beta_{80}$ - defined in Eq.~(\ref{eq:radius}), using the Imaginary Water Flow technique.
All details regarding the methodology of searching for the right saddles
with the exact specification of the deformation spaces used, can be found in Ref.~\cite{Jach2017}.
Finally, we want to emphasize, that recently we have systematically determined inner and outer fission barrier heights
for 75 actinides, within the range from actinium to californium, including
odd-$A$ and odd-odd systems, for which experimental estimates were accessible \cite{actinides}.
A statistical comparison of our inner and outer fission barrier heights with available
experimental estimates gives the average discrepancy and the rms deviation
not larger than 0.82 MeV and 0.94 MeV, respectively.
This allows us to have some confidence in the macroscopic-microscopic model used here.
Significant differences in the fission barriers obtained in various modern nuclear models were noticed in Ref.~\cite{bro2014}.
A broad discussion of the problems arising from this can be found in Refs.~\cite{ Baran2015, Jach2017}.

 Owing to the dependence of the shell effects on nuclear excitation,
the value of shell correction effectively depends on the excitation energy
with the damping parameter $E_d=25$ MeV.
In comparison to Refs.~\cite{SWCK,SWCK2,SWCK3}, which are based for even-even nuclei on the same mass table, we employ the
equivalent method to calculate the survival probability \cite{model2,model3,paper1} taking into account the shell effect damping
in the potential energy surface and asymptotic level-density parameter $a$.
However, we would like to emphasise that for odd nuclei in Refs.~\cite{SWCK,SWCK2,SWCK3}
the pairing was treated in different way compared to nuclear input data used here.
Namely, the predictions of the Fusion-by-Diffusion model~\cite{SWCK,SWCK2,SWCK3} for the synthesis cross sections of 114–120 elements
were based on the macroscopic-microscopic properties calculated within the quasiparticle method in pairing channel.
The ground states and consequently fission barrier heights for other nuclei were calculated
separately by adding the energy of the odd particle
occupying a single-particle state. This quasiparticle energy $E_{qp}$
in the superconducting state takes a simple form:
$E_{qp} = \sqrt{(\varepsilon_{qp} -\lambda)^{2} + \Delta^{2}}$, where $\varepsilon_{qp} $ is the energy of the odd nucleon
in the quasiparticle state, $\lambda$ is the Fermi energy and $\Delta$ is the pairing gap
energy.
In this scenario of fission barriers calculation the energy $E_{qp}$ was added at every grid point
as well as at every minimisation step in the gradient procedure used for the ground states.
So, the calculations of masses and $B_{f}$ have been performed without blocking of any state in the calculations
within the Fusion-by-Diffusion model~\cite{SWCK,SWCK2,SWCK3}
but with using the BSC-quasiparticle method.

In the DNS model used here the damping parameter should be larger than in Refs.~\cite{SWCK,SWCK2,SWCK3}.
With the expression $a_n=a=A/10$ MeV$^{-1}$ for the asymptotic level-density parameter
for neutron ($A$ is the mass number of the CN) we obtain almost the same values as those used in Ref.~\cite{CSWW}
and found microscopically  in Ref.~\cite{AZAM}.
The level-density parameters for   fission, proton-emission, and $\alpha-$emission channels
are taken as $a_f=0.98 a$, $a_p=0.96 a$, and $a_\alpha=1.15 a$, respectively. The ratio between $a$ and $a_f$ is closed to that
found in  Ref.~\cite{AZAM}. Here, we set these parameters for all reactions considered.
Because the shell corrections at the ground state are larger with the mass table \cite{moller}, in Refs.~\cite{model2,model3}
the larger values of $a_f=1.03 a$ were used. Other parameters in Refs.~\cite{model2,model3} were set the same as in this paper.
So, taking other mass table for the properties of SHN, we change only the ratio $a_f/a$.

For the calculation of the Coulomb barrier, we use  the expression
\begin{equation}
V_j=\frac{(Z-z_j)z_je^2}{r_j[(A-m_j)^{1/3}+m_j^{1/3}]},
\label{coulomb}
\end{equation}
where $z_j$ ($m_j$) are   the charge  (mass)
numbers of the charged particle (proton or $\alpha$-particle)
and $r_j$ is a constant.
The charge $Z$ (mass $A$) number corresponds to the CN.
There are different theoretical estimations of $r_j$ \cite{BT,rp}.
In the case of $\alpha-$emission, $r_\alpha$ varies from
1.3 to 1.78 fm. We obtain $r_\alpha$ from  the energy of the
DNS formed by the daughter nucleus and $\alpha$-particle.  We
calculate the Coulomb barrier in the interaction potential between the
$\alpha$-particle and the daughter nucleus \cite{poten1}, and find
the value of $r_\alpha$  from Eq. (\ref{coulomb}).
For different nuclei
considered, we obtained $r_\alpha=1.57$ fm using this method. Thus,
in the calculations of $V_\alpha$ we set $r_\alpha=1.57$ fm  for nuclei considered.
The parameter $r_p$
for the Coulomb barrier for proton emission is taken as $r_p$=1.7 fm
from Refs. \cite{charge,rp}.
With these values of $r_\alpha$ and $r_p$ Eq.~(\ref{coulomb}) results in
$V_\alpha$ and $V_p$ which are about 2.5 and 1.5 MeV (Table \ref{barriers}), respectively, larger than those
used in Refs.~\cite{SWCK,SWCK2,SWCK3}. As shown in Refs.~\cite{SWCK,SWCK2,SWCK3}, the increase of
$V_\alpha$ and $V_p$ by 4 MeV leads to about one order of magnitude smaller $\sigma_s$
in the $\alpha xn$ and $pxn$ evaporation channels. So, the difference of our $r_\alpha$ and $r_p$ from
those in Refs.~\cite{SWCK,SWCK2,SWCK3} could create 2--4 times difference in the values of $\sigma_s$.
In Refs.~\cite{model2,model3}, the same values of $V_\alpha$ and $V_p$ were used as in this paper.
As seen in Table \ref{barriers}, the values of energy thresholds for protons and alpha-particles obtained with the mass
table \cite{moller} deviate within 2.5 MeV from those calculated with the mass table \cite{MKowal}.

 As  found, the values of $\sigma_s$ near the maxima of excitation functions are almost insensitive to the
reasonable variations of the parameters used,  but  far from the maxima they change
up to one order of magnitude. Therefore,
the results obtained in this paper have quite a small uncertainty
near the maxima of excitation functions which are
important to get the maximum yield of certain nucleus in the experiments.
We estimate the uncertainty of our calculations of $\sigma_s$ within a factor of 2--4.
Our
model was well tested in Ref.~\cite{paper1} for many reactions in which the excitation
functions of transfermium nuclei produced in the charged particle evaporation channels
have been measured.

\begin{figure}
\includegraphics[width=0.45\textwidth,clip]{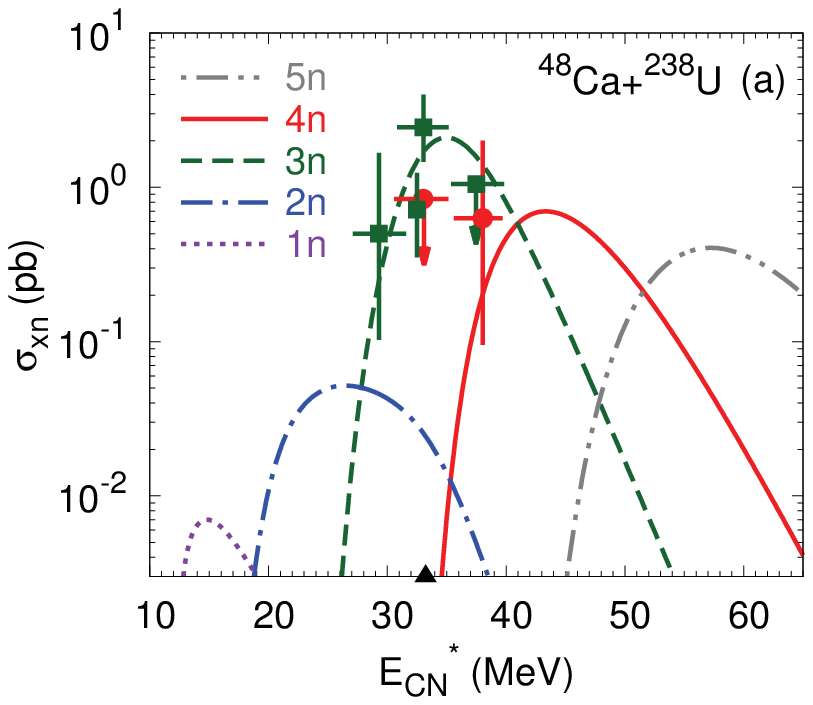}
\includegraphics[width=0.45\textwidth,clip]{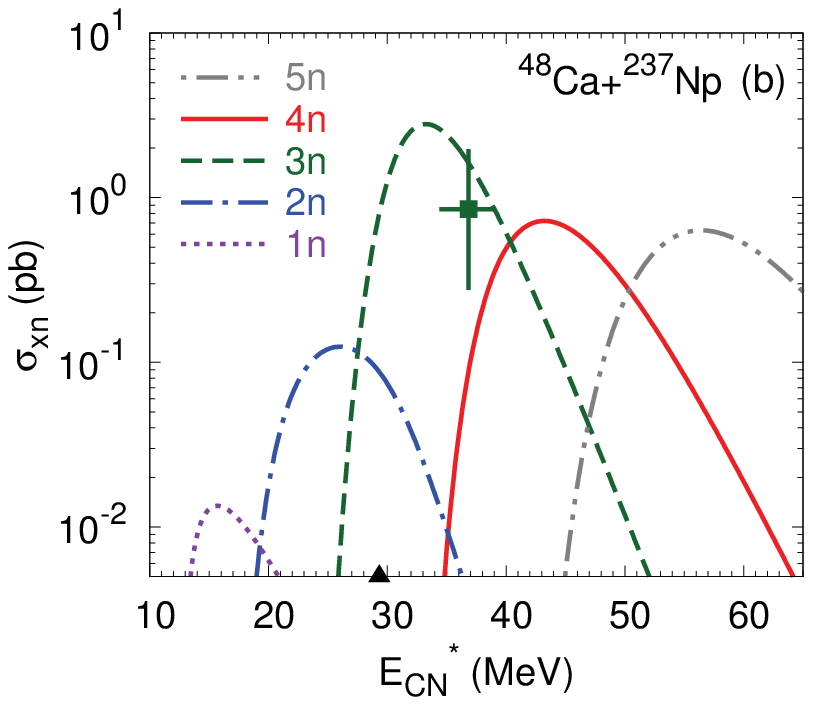}
\includegraphics[width=0.45\textwidth,clip]{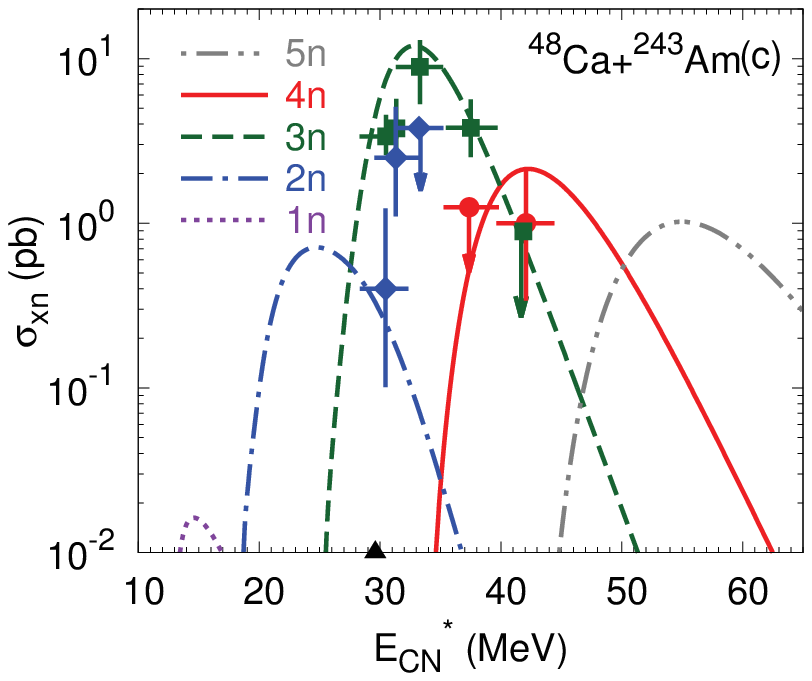}
\includegraphics[width=0.45\textwidth,clip]{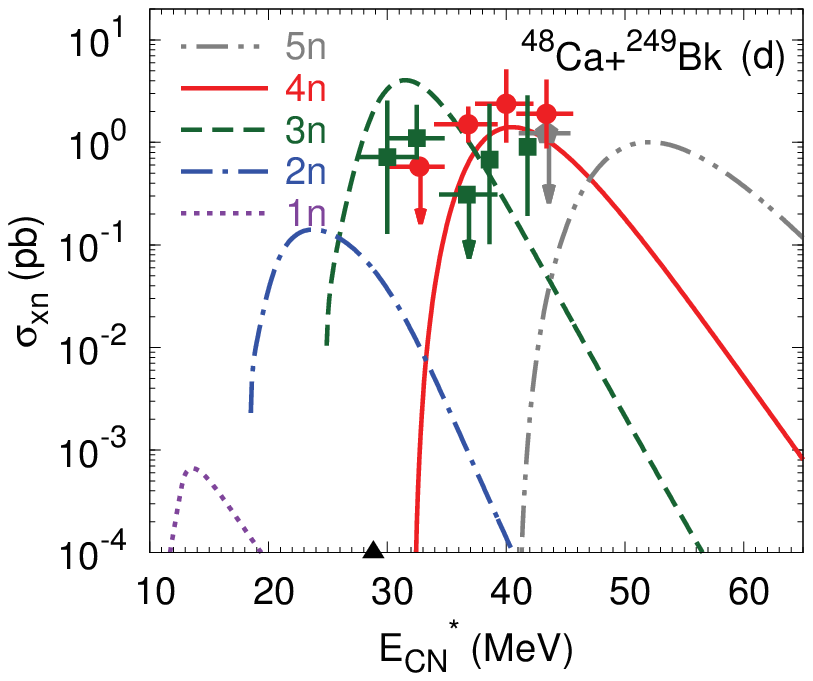}
\caption{
The measured (symbols) and calculated (lines) excitation functions for  $xn-$evaporation channels ($x=1-5$) of the indicated complete fusion  reactions.
The mass table of
Ref.~\cite{MKowal} is used  in the calculations.
The  black  triangles at energy axis indicate the excitation energy
$E_{CN}^*$ 
of the CN at bombarding energy corresponding to the Coulomb barrier 
for the sphere-side orientation.
The blue diamonds, green squares,
red circles, and gray pentagons represent the experimental data \cite{Og1n} with error bars for $2n-$,
$3n-$, $4n-$, and $5n-$evaporation  channels, respectively. The vertical lines with
arrow indicate the upper limits of evaporation residue cross sections.
}
\label{caactn}
\end{figure}
%
%

%
\begin{figure}
\includegraphics[width=0.45\textwidth,clip]{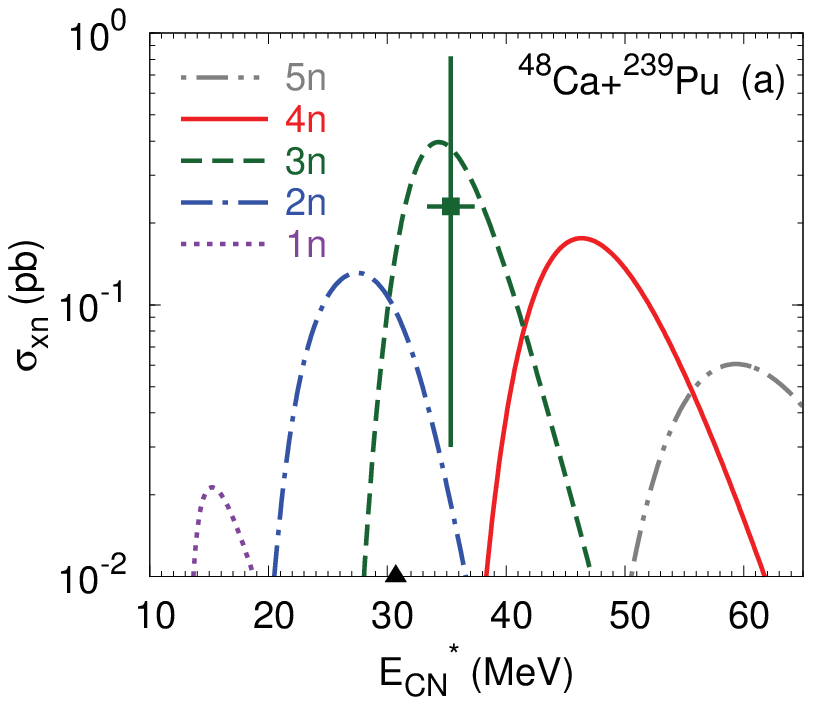}
\includegraphics[width=0.45\textwidth,clip]{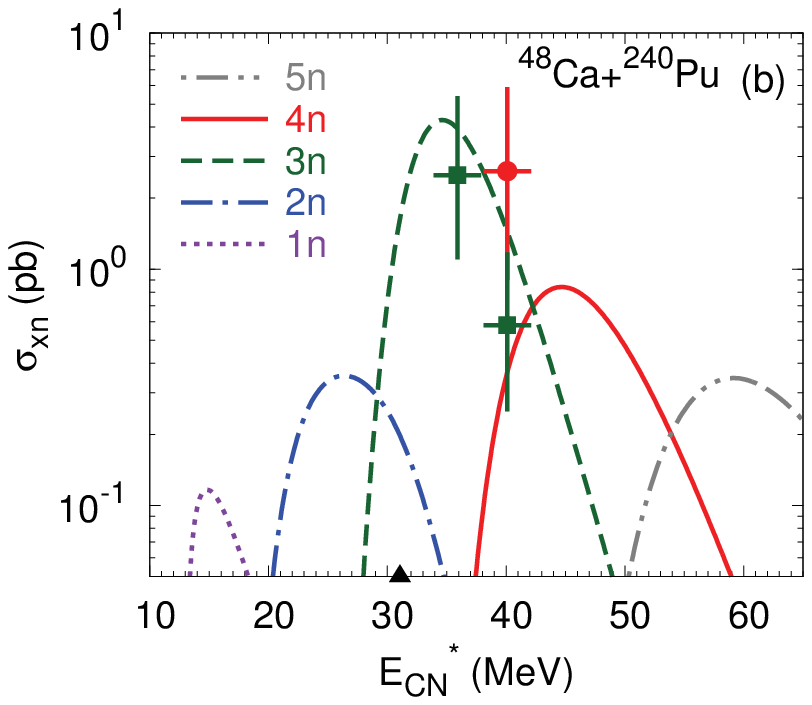}
\includegraphics[width=0.45\textwidth,clip]{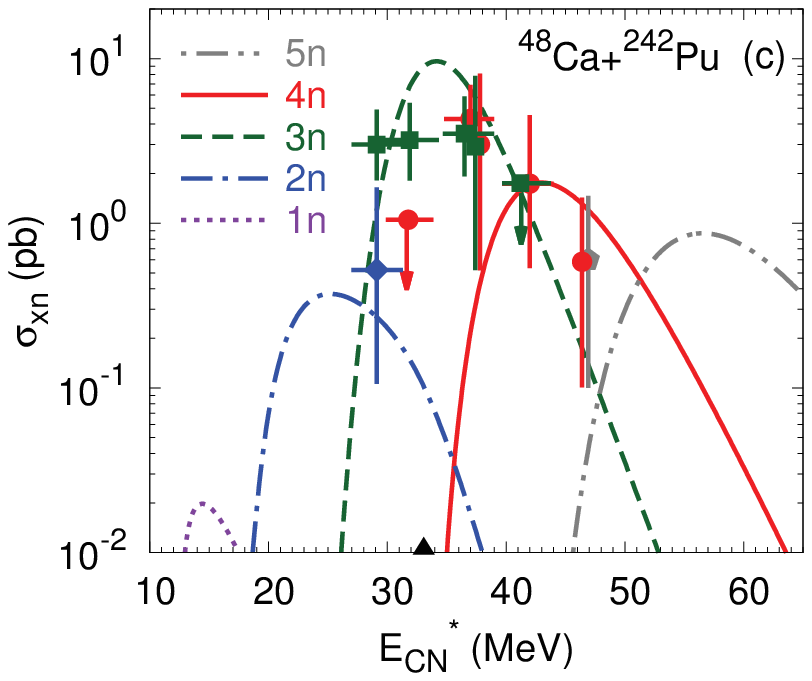}
\includegraphics[width=0.45\textwidth,clip]{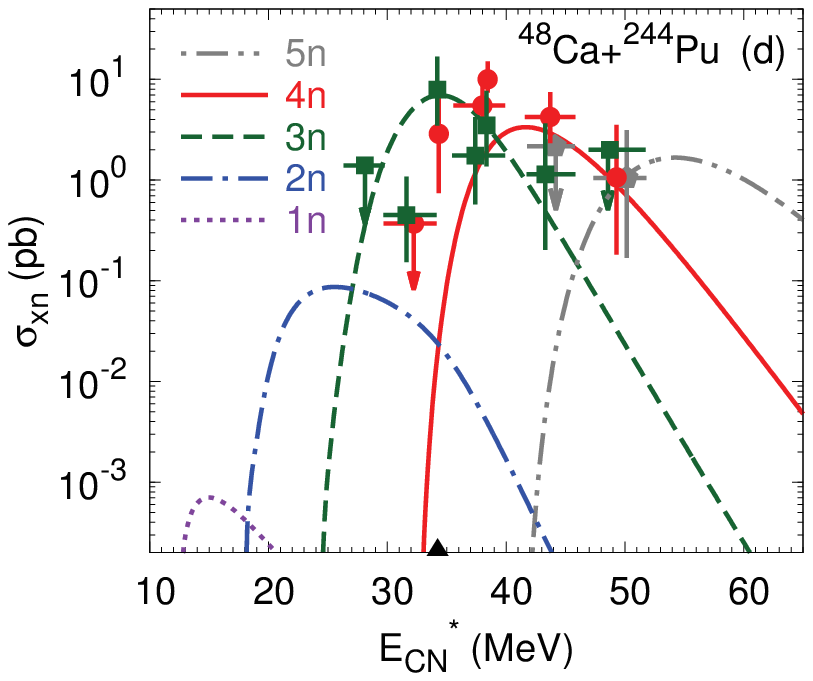}
\caption{
The same as in Fig.~\ref{caactn}, but for other indicated  complete fusion  reactions.
}
\label{caactn2}
\end{figure}
\begin{figure}
\includegraphics[width=0.45\textwidth,clip]{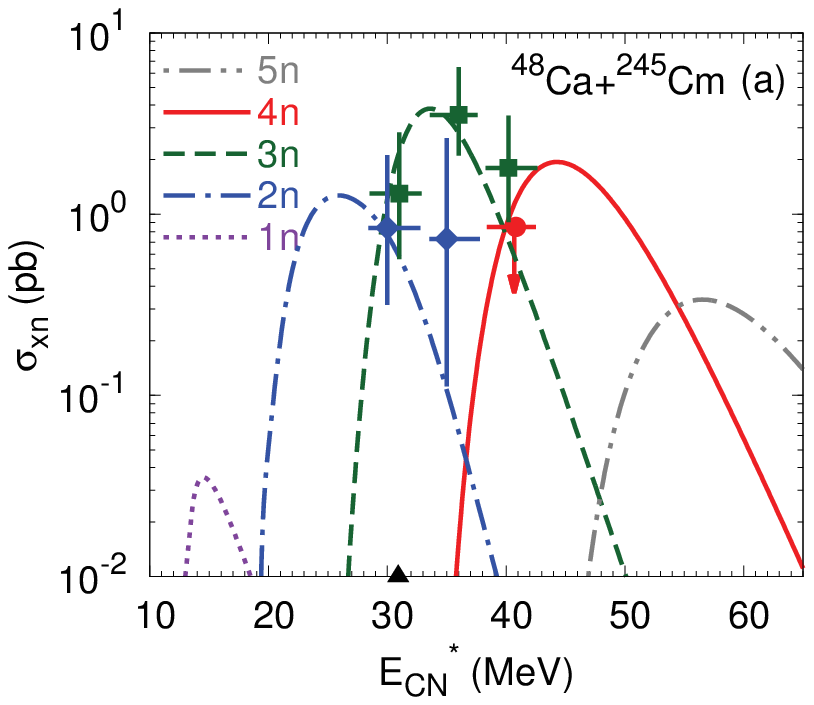}
\includegraphics[width=0.45\textwidth,clip]{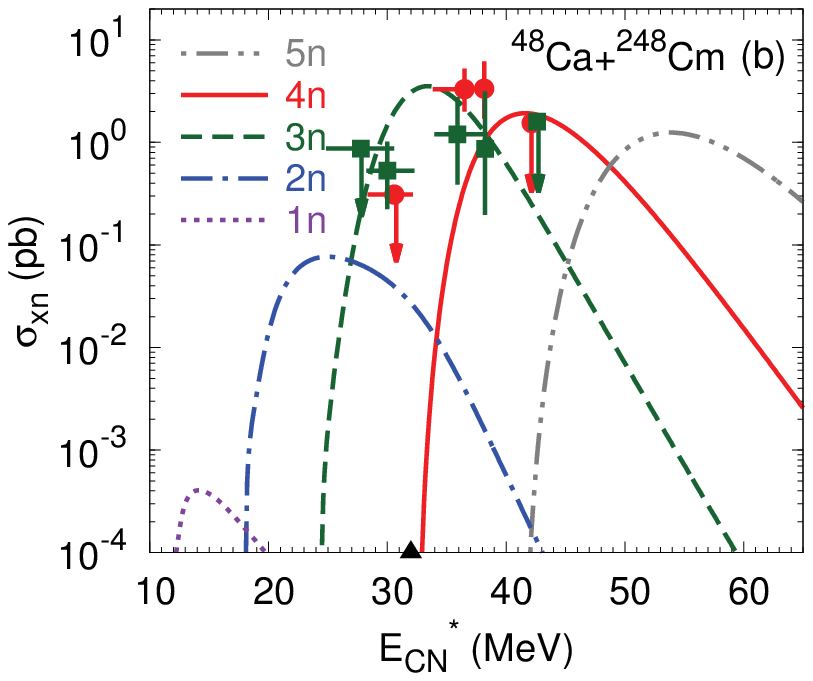}
\includegraphics[width=0.45\textwidth,clip]{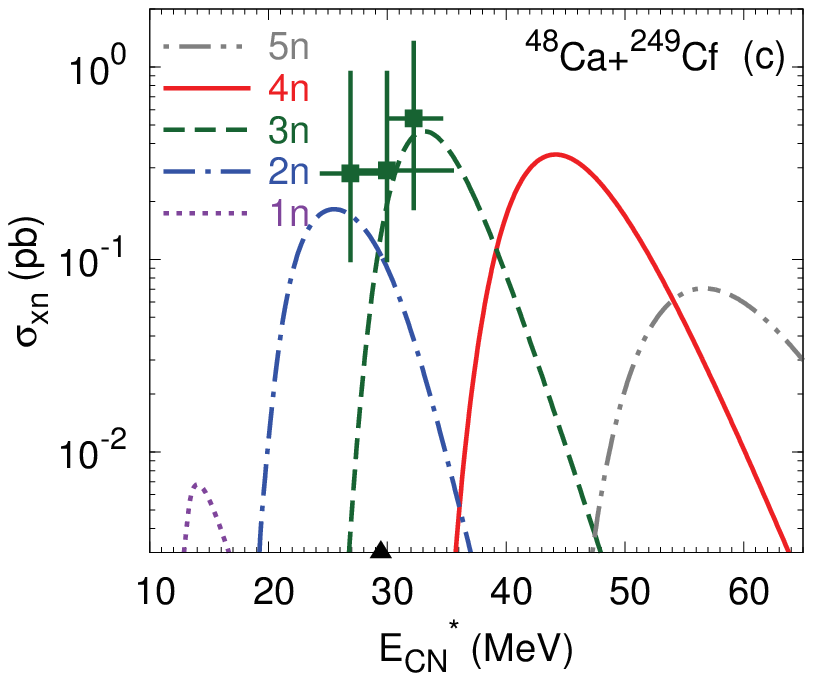}
\includegraphics[width=0.45\textwidth,clip]{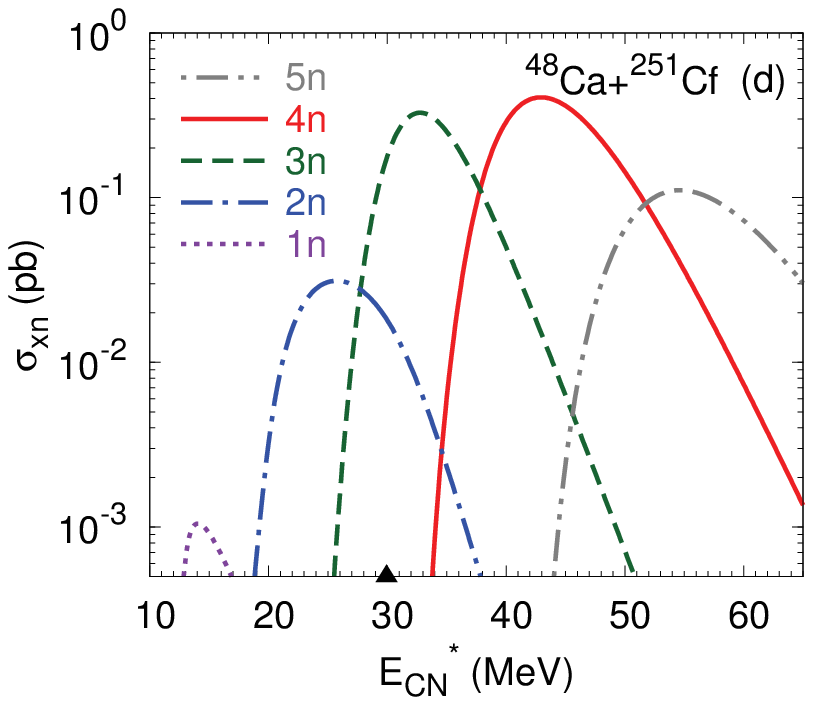}
\caption{
The same as in Fig.~\ref{caactn}, but for other indicated  complete fusion  reactions.
}
\label{caactn3}
\end{figure}
\begin{figure}
\includegraphics[width=0.45\textwidth,clip]{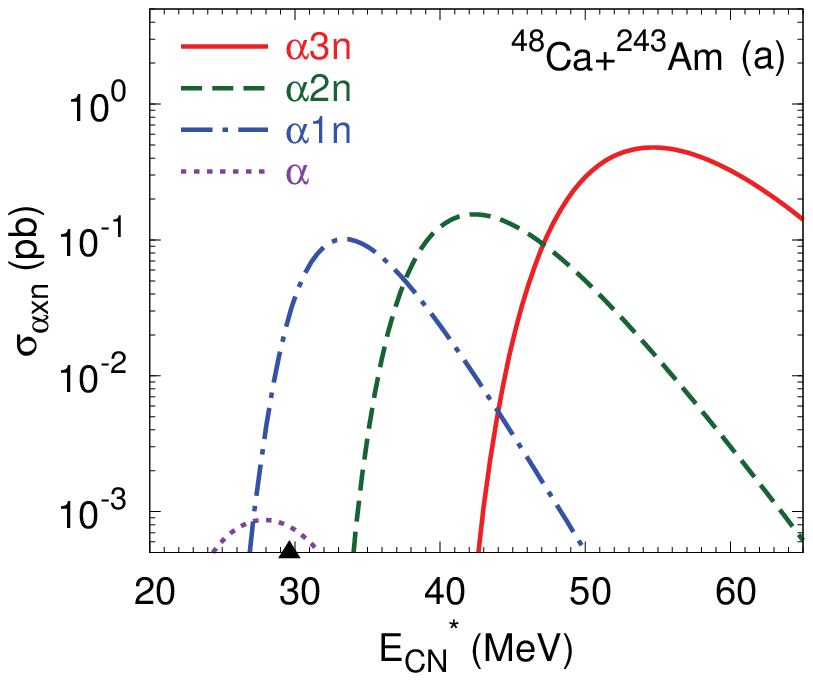}
\includegraphics[width=0.45\textwidth,clip]{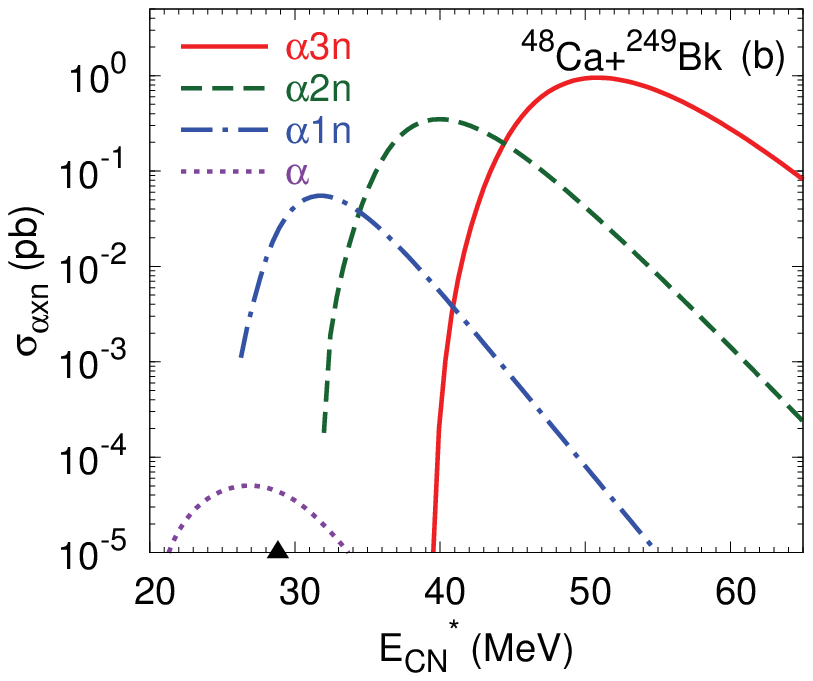}
\includegraphics[width=0.45\textwidth,clip]{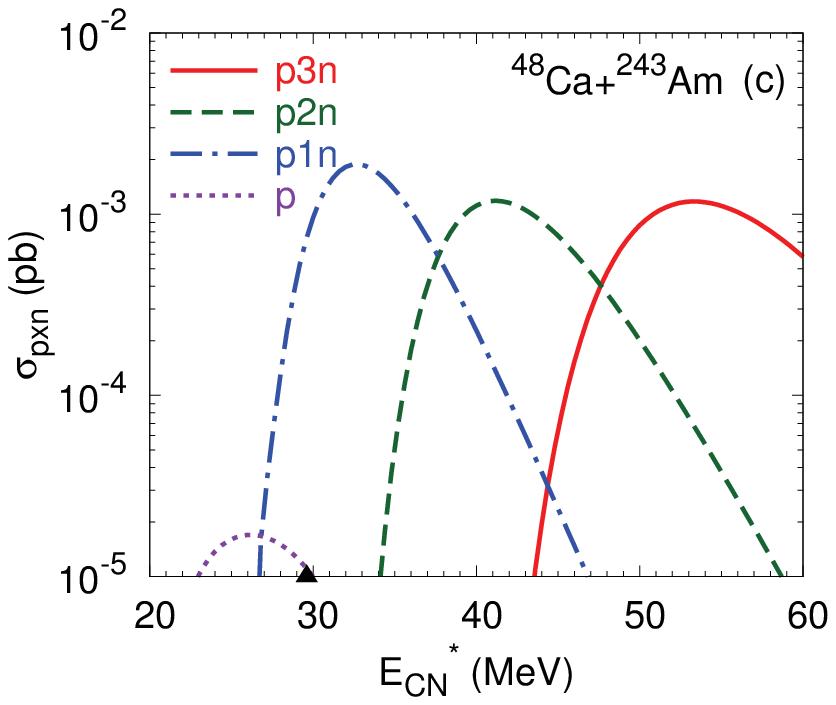}
\includegraphics[width=0.45\textwidth,clip]{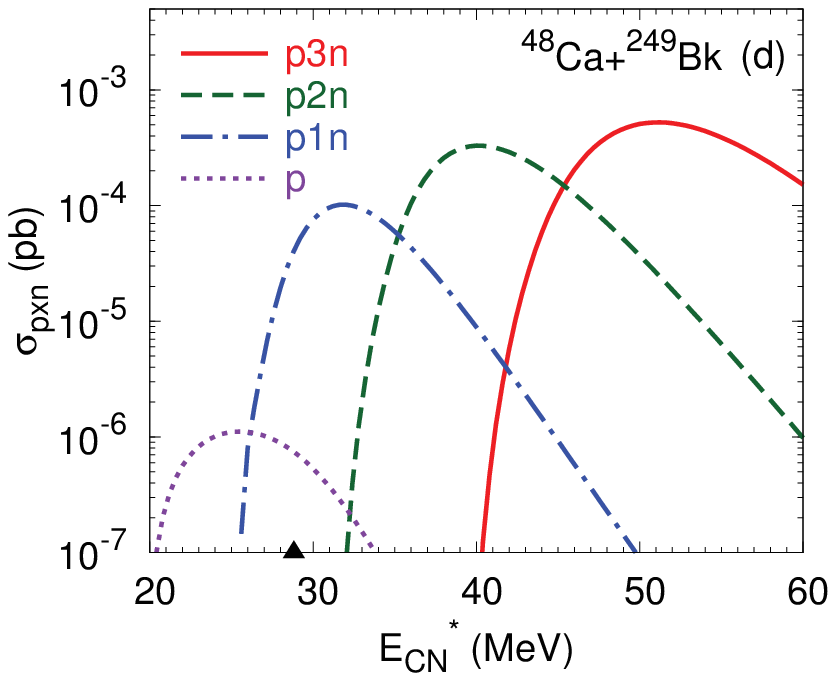}
\caption{
The same as in Fig.~\ref{caactn}, but for  $\alpha xn-$evaporation channels ($x=0-3$) of the indicated complete fusion  reactions.
}
\label{caactn4}
\end{figure}
\begin{figure}
\includegraphics[width=0.45\textwidth,clip]{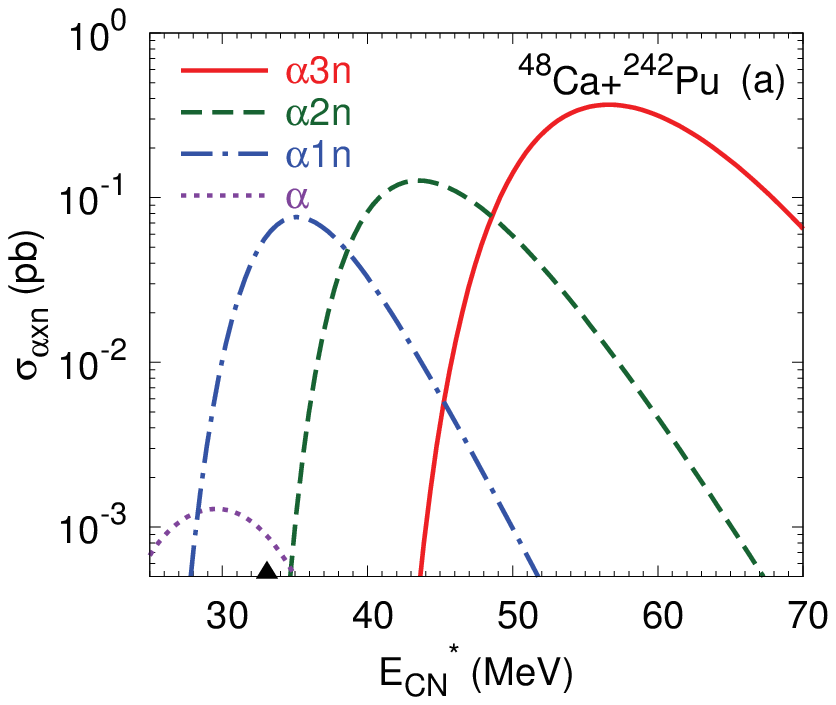}
\includegraphics[width=0.45\textwidth,clip]{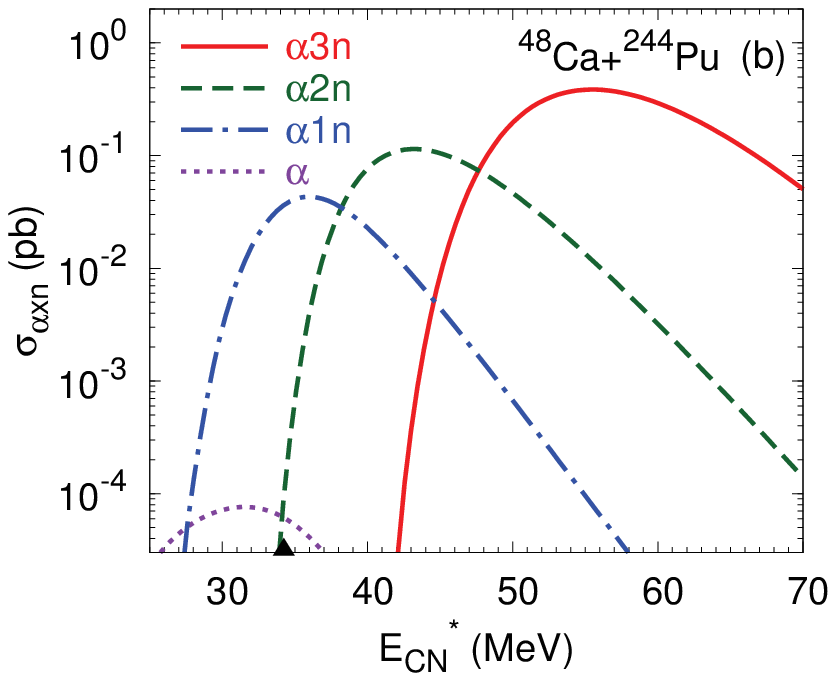}
\includegraphics[width=0.45\textwidth,clip]{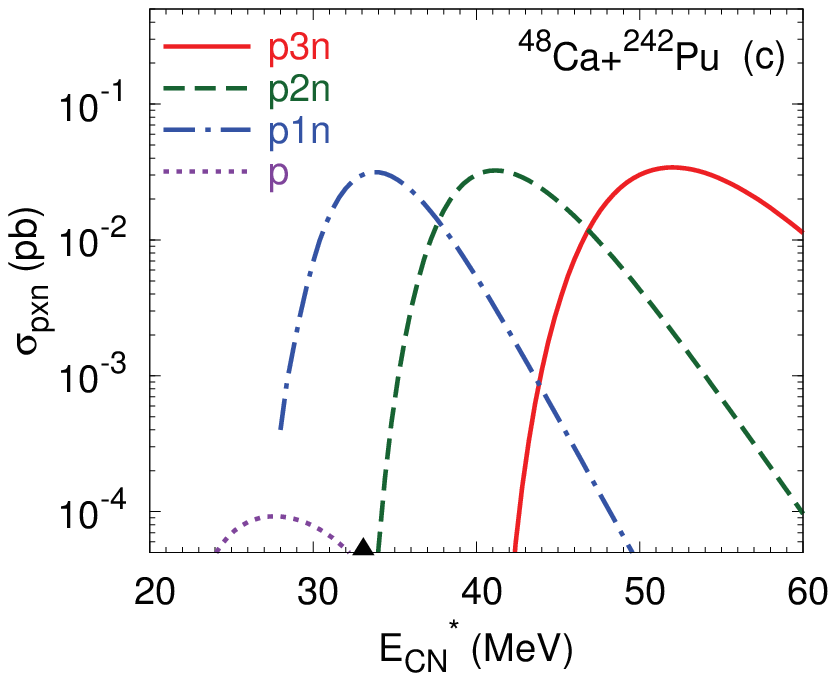}
\includegraphics[width=0.45\textwidth,clip]{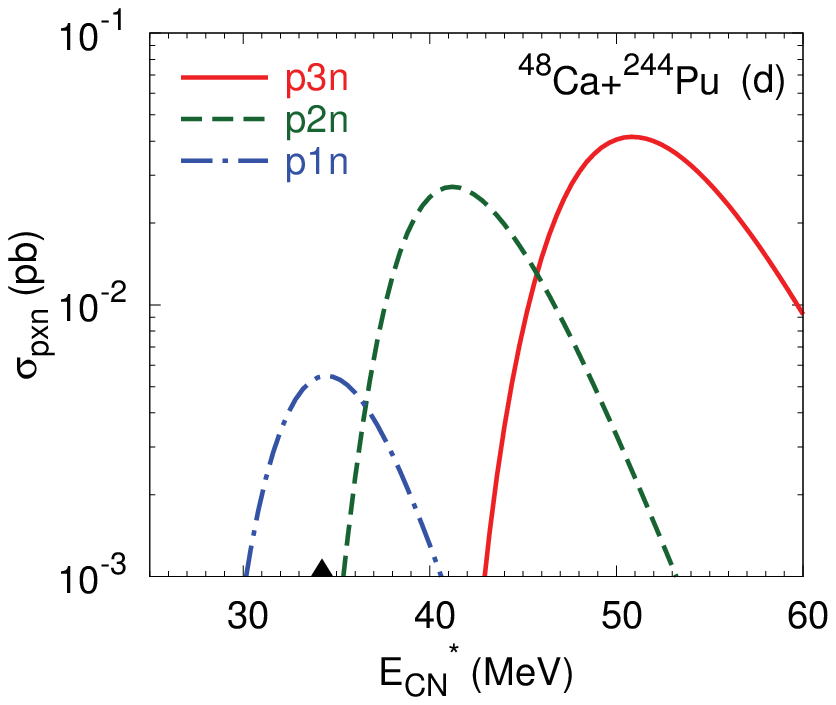}
\caption{
The same as in Fig.~\ref{caactn}, but for  $\alpha xn-$ and $pxn-$evaporation channels ($x=0-3$) of the indicated complete fusion  reactions.
}
\label{caactn5}
\end{figure}
\begin{figure}
\includegraphics[width=0.45\textwidth,clip]{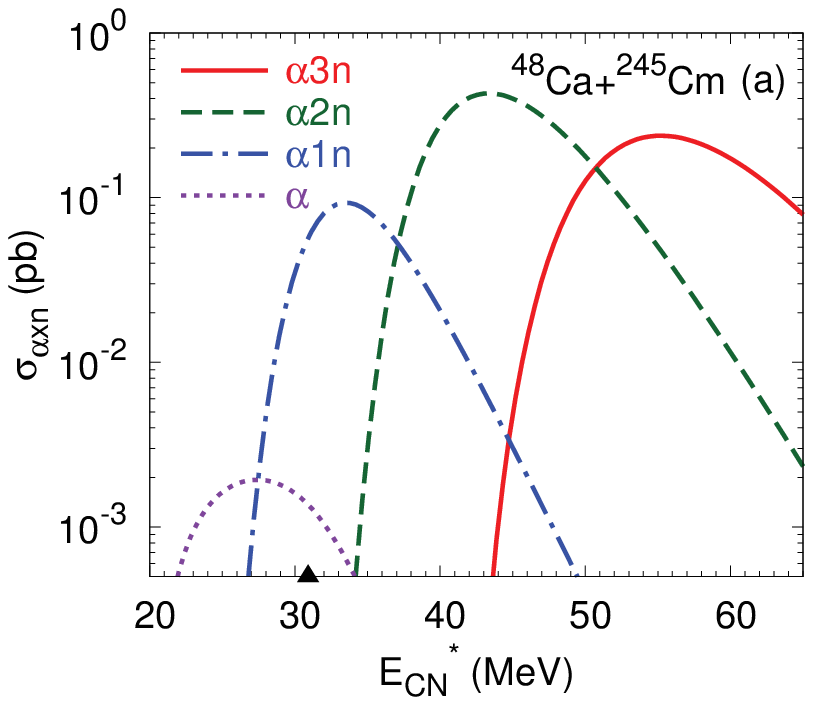}
\includegraphics[width=0.45\textwidth,clip]{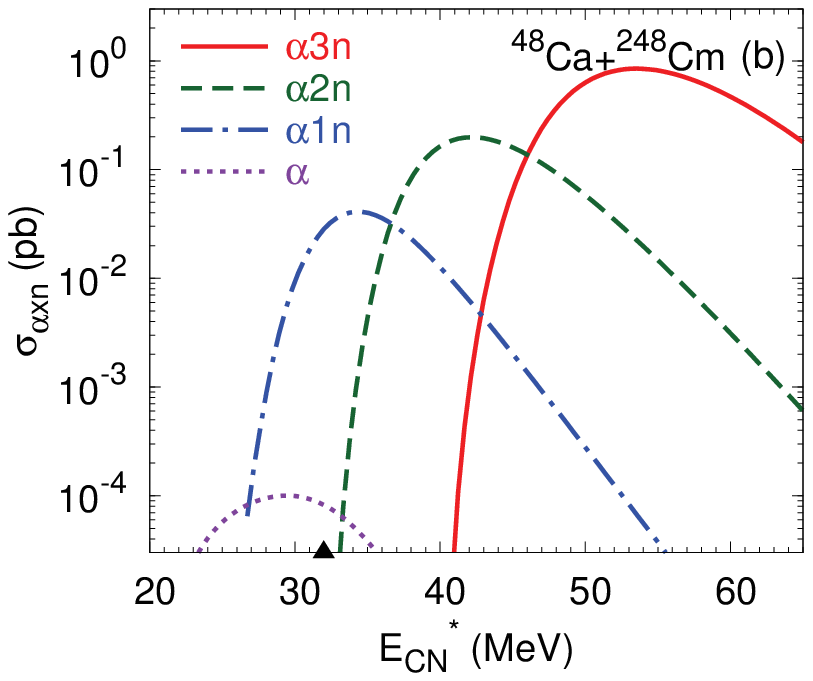}
\includegraphics[width=0.45\textwidth,clip]{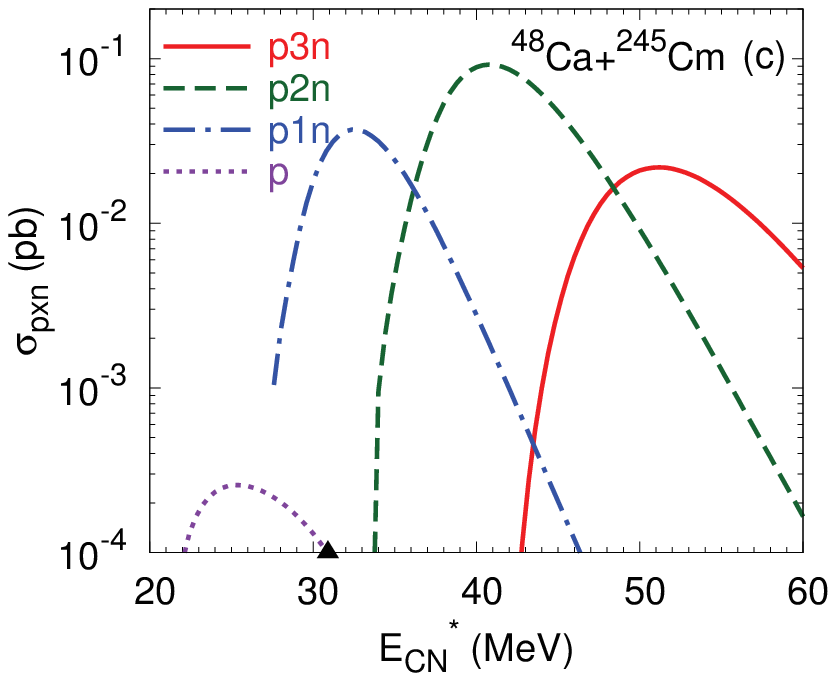}
\includegraphics[width=0.45\textwidth,clip]{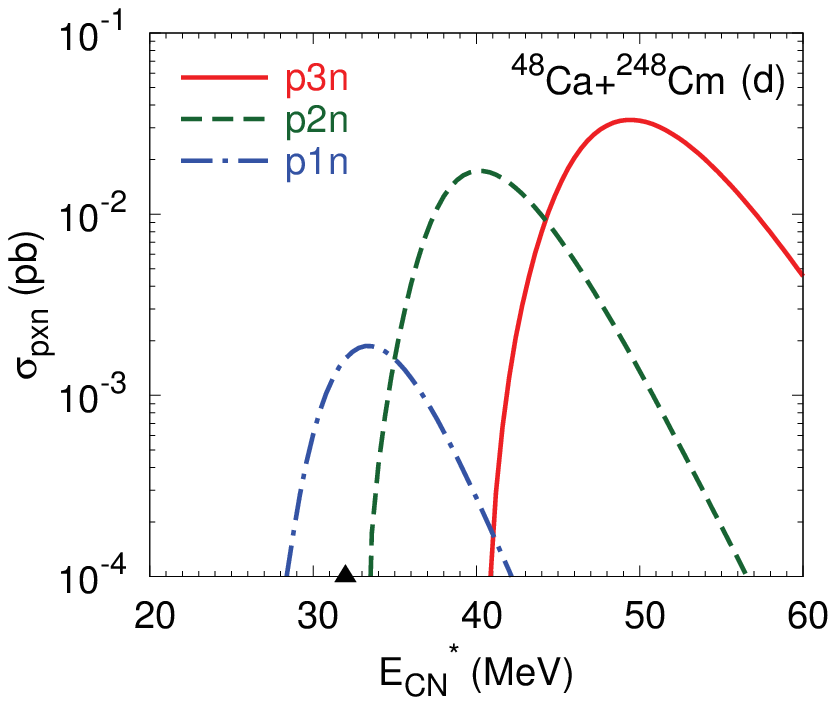}
\caption{
The same as in Fig.~\ref{caactn}, but for  $\alpha xn-$ and $pxn-$evaporation channels ($x=0-3$) of the indicated complete fusion  reactions.
}
\label{caactn6}
\end{figure}
\begin{figure}
\includegraphics[width=0.45\textwidth,clip]{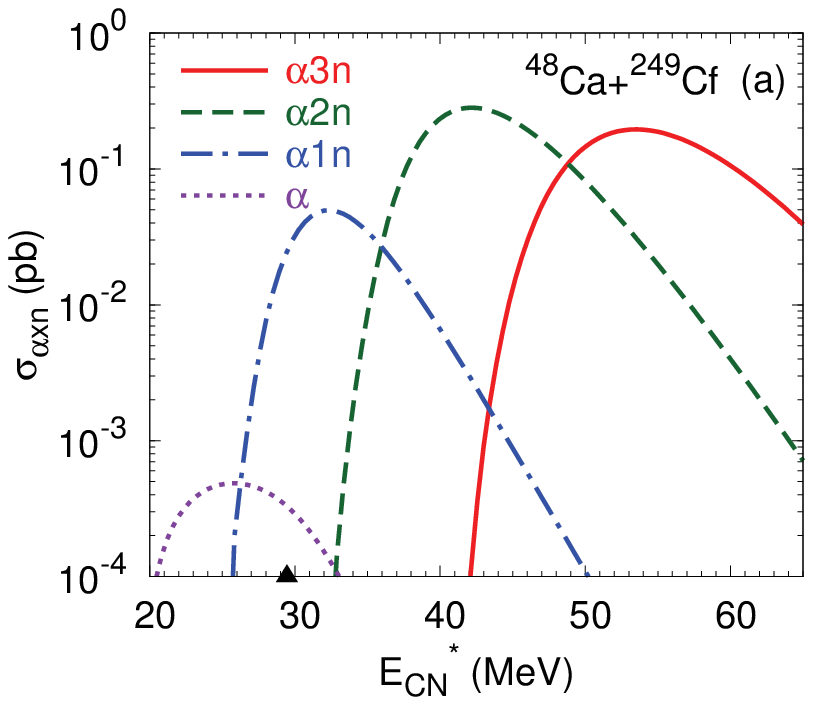}
\includegraphics[width=0.45\textwidth,clip]{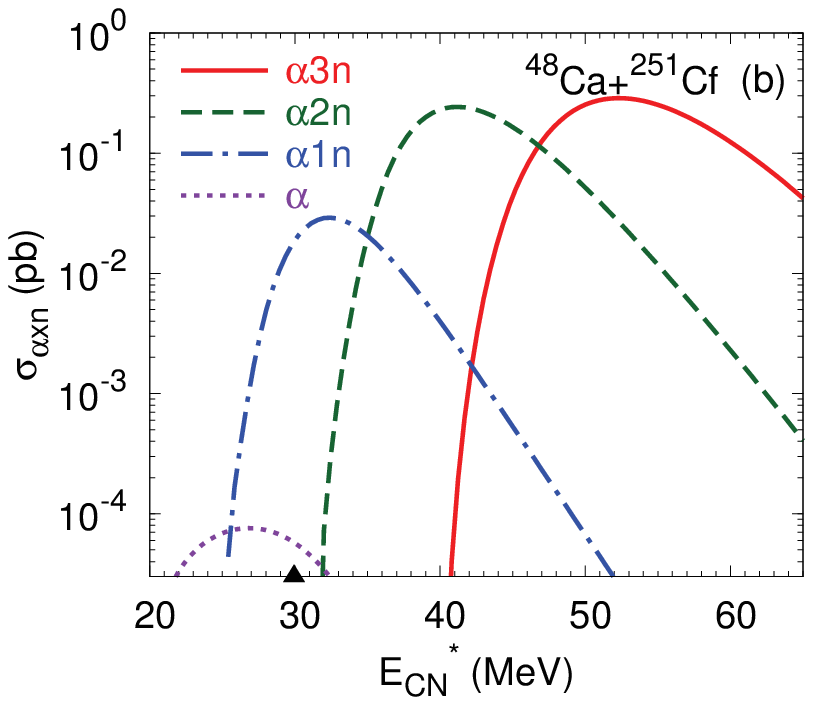}
\includegraphics[width=0.45\textwidth,clip]{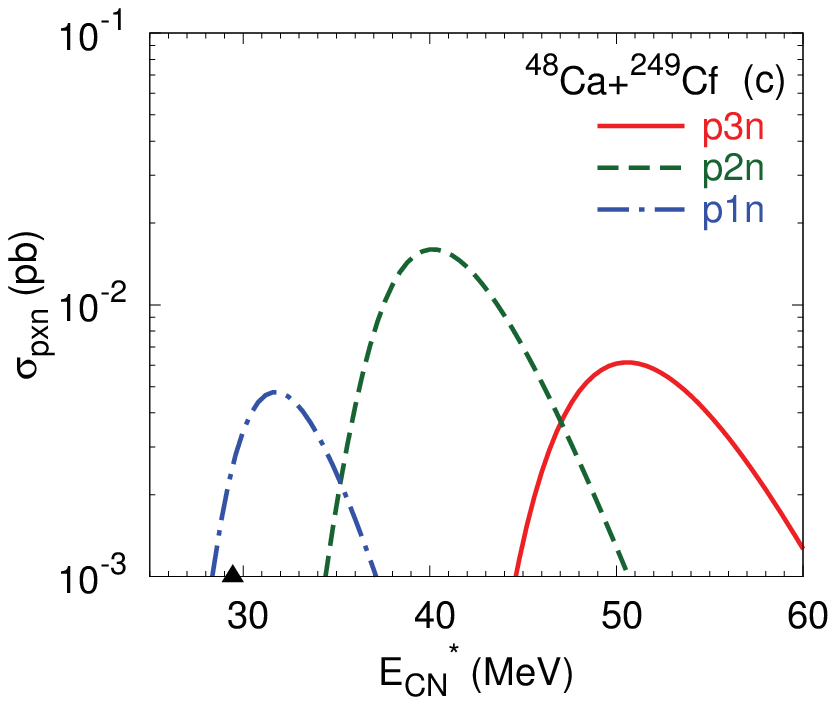}
\includegraphics[width=0.45\textwidth,clip]{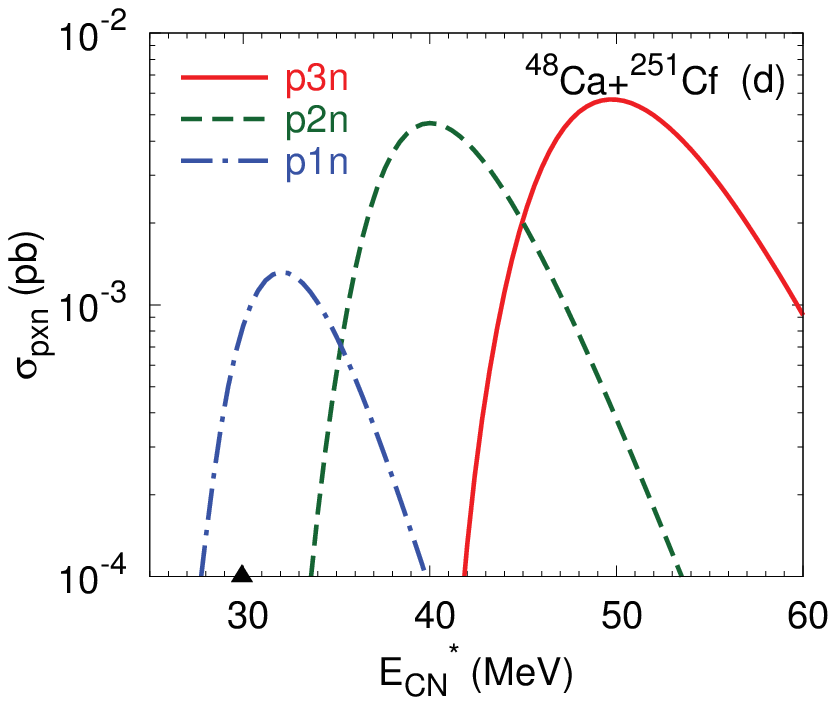}
\caption{
The same as in Fig.~\ref{caactn}, but for  $\alpha xn-$ and $pxn-$evaporation channels ($x=0-3$) of the indicated complete fusion  reactions.
}
\label{caactn7}
\end{figure}
\section{Calculated results}
In Fig.~\ref{masses}, our results for nuclear masses (top panels) and calculated from them $Q_{\alpha}$-values (bottom panels)
for SHN considered in this article are shown. As one can see, the available alpha-decay energy
measurements are perfectly reproduced.
Only in the case of Cn and Nh nuclei with smaller number of neutrons our results slightly overestimate the experimental data.
The exact values in some of the most important cases here are summarized in the Table \ref{tab:Spect_L}.
Let us emphasize that only ground-state-to-ground-state alpha transition were calculated.
Apparent $Q_{\alpha}$ values taking the parent ground-state configuration in odd and odd-odd systems
as the final state in daughter were not considered.
This may be the reason for the overestimation in a few cases,
as especially in odd nuclei the decay may occur to excited states of the daughter nucleus,
which shortens the alpha transition lines.

In Fig.~\ref{fisbar} we provide a comparison of our calculated fission barriers with
the available experimental estimates based on the observed evaporation residue production probabilities \cite{ITKIS}.
As in the case of Fig.~\ref{masses}, the isotopes: Fl, Cn, Lv and Og are shown on the left-hand side while
Nh, Mc, Ts on the right-hand side.
As seen in Fig.~\ref{fisbar}, our
calculated  fission barriers  $B_f$ are in quite good agreement with the experimental (empirical) estimates \cite{ITKIS}
marked by crosses.
For completeness, we show the neutron separation energies $B_n$  in the middle part of Fig.~\ref{fisbar}.
The strong staggering is clearly visible.
As mentioned before, the height of this threshold in relation to the fission barrier is crucial
for determining which process wins the competition in nuclear decay, fission or neutron emission.
In Fig.~\ref{fisbar} (bottom panels), the differences between these two key parameters controlling survival probability are also shown.

With the nuclear properties from Refs.~\cite{MKowal, Jach2014, Jach2017}
the calculated excitation functions for $xn$ evaporation channels are presented in Figs.~\ref{caactn}-\ref{caactn3}
for the complete fusion reactions $^{48}$Ca+$^{238}$U,$^{237}$Np,$^{243}$Am,$^{249}$Bk,$^{239,240,242,244}$Pu,$^{245,248}$Cm,$^{249,251}$Cf.
In Ref.~\cite{model3} and here, the same model is used to calculate the evaporation residue cross sections. So, the comparison with
the results of Ref.~\cite{model3} reflects the difference in the predicted properties of SHN.
In comparison to Ref.~\cite{model3} the bombarding energies corresponding to the Coulomb barriers for the sphere-side
orientations lead to 2--3 MeV smaller $E_{CN}$. As a result, the maxima of excitation functions $\sigma_{xn}(E_{CN})$
are slightly shifted to higher $E_{CN}$ in Figs.~\ref{caactn}-\ref{caactn3}. If in Ref.~\cite{model3} $\sigma_{4n}>\sigma_{3n}$
in the maxima of excitation functions for the reactions $^{48}$Ca+$^{242,244}$Pu,$^{245,248}$Cm, the present
calculations with the data of Ref.~\cite{MKowal} result in $\sigma_{4n}<\sigma_{3n}$ and larger $\sigma_{3n}/\sigma_{2n}$.
Though in the reactions $^{48}$Ca+$^{238}$U,$^{237}$Np,$^{243}$Am,$^{249}$Bk the maximum production cross sections are
expected in the $3n$ evaporation channel independently on the mass table, in Fig.~\ref{caactn} the ratios $\sigma_{3n}/\sigma_{4n}$
are about 2 times smaller than those in Ref.~\cite{model3}. The mass table \cite{MKowal} leads to close maxima of $\sigma_{3n}$ and $\sigma_{4n}$,
relatively large $\sigma_{5n}$ and smaller $\sigma_{2n}$ in most reactions.

The maximum cross sections in the $2n$-evaporation channel were
found to be within factor of 10 smaller than the cross sections at the maxima of excitation
functions of the $3n-$ or $4n-$evaporation channels. The cross sections in $1n$ evaporation channel could be of interest for
the experimental study if they are larger than 5 fb.
 Thus, employing reactions in the $1n-$ and $2n-$evaporation channels,
one can directly produce heaviest isotopes closer
to the center of "the island of stability":
$^{284,285}$Cn,
$^{283,284}$Nh,
$^{294}$Lv,
$^{295}$Ts,
and
$^{295-297}$Og.
Many of them were only produced as daughters in the $\alpha-$decay chains.
The isotopes
$^{295}$Ts,
and
$^{295-297}$Og
are presently
unknown.

The comparison of the results in Figs.~\ref{caactn}-\ref{caactn3} with those in Refs.~\cite{SWCK2,SWCK3} based on the same
mass table allows us to stress the difference of the fusion models used. In spite of the difference of the fusion models,
the predicted values of cross sections are rather close for most reactions. While  $\sigma_{4n}>\sigma_{3n}$ for the $^{48}$Ca+$^{249}$Cf reaction
in Ref.~\cite{SWCK2}, in Fig.~\ref{caactn3} we obtain $\sigma_{4n}<\sigma_{3n}$. For the $^{48}$Ca+$^{249}$Bk, we obtain smaller ratio
$\sigma_{4n}/\sigma_{5n}$ and larger $\sigma_{3n}/\sigma_{4n}$ than those in Ref.~\cite{SWCK3}.

The calculated excitation functions for the channels with evaporation of charged particle are presented in Figs.~\ref{caactn4}-\ref{caactn7}.
While in Ref.~\cite{model2} $\sigma_{\alpha 2n}>\sigma_{\alpha 3n}$ and $\sigma_{p 2n}>\sigma_{p 3n}$ with mass table \cite{moller} in most reactions,
we obtain rather close cross sections in $\alpha 2n$ and $\alpha 3n$, $p2n$ and $p3n$ evaporation channels due to slightly smaller
neutron separation energies in the mass table \cite{MKowal, Jach2014, Jach2017}.
Because the same mass table is used (with a reservation regarding the different treatment of odd particles) in Ref.~\cite{SWCK},
the predicted cross sections are similar there to those in Figs.~\ref{caactn4}-\ref{caactn7}.
However, stronger dependence of fusion probability $P_{CN}$ on energy leads to relative increase of
the $\alpha 3n$ and $p3n$ evaporation channels
in  Figs.~\ref{caactn4}-\ref{caactn7}. The relatively smaller yields in the $\alpha 1n$ and $p1n$ evaporation channels
are due to the same reason.

The production cross sections  of almost all of these SHN  in the $xn$-evaporation channels
are comparable or even larger than those in the charged particle evaporation channels.
The production cross sections of
 heaviest isotopes
$^{287-290}$Nh, $^{291-293}$Mc, and $^{295-296}$Ts
($^{286,287}$Cn, $^{286}$Nh, $^{290,291}$Fl, $^{291,292}$Mc, and $^{294}$Lv)
in the $pxn$-channels ($\alpha xn$-channels)
of the $^{48}$Ca-induced fusion reactions
were predicted:
about 10--200 fb in the $pxn-$evaporation channels (about 50--500 fb
in the $\alpha xn-$evaporation  channels).

\section{Summary}
%

For the $^{48}$Ca-induced actinide-based  complete  fusion   reactions,
the excitation functions for the  production of the SHN with charge numbers
112--118 were calculated in  $xn-$, $\alpha xn$-, and $pxn$-evaporation channels
using the predictions of SHN properties from Ref. \cite{MKowal, Jach2014, Jach2017}.

As it turns out, in modeling of reactions leading to the SHN,
the use of a consistent i.e., coming from one source, set of nuclear data input plays a fairly important role.
In the presented article, all used nuclear properties of the ground states and saddle points
were calculated within the multidimensional macroscopic-microscopic approach with blocking technique for odd nuclei.

Obtained agreement of cross-sections for the reactions in $3n$ channel is excellent.
Excitation functions are only slightly shifted towards higher energies compared to the experiment when four neutrons are emitted in the cascade.
Only for the reactions; $^{48}$Ca+$^{240}$Pu and $^{48}$Ca+$^{242}$Pu,
the resulting cross-sections are underestimated - but less than one order of magnitude.

The use of the charged particle evaporation channels allows us to increase
the mass number of heaviest isotopes of nuclei with $Z$= 113, 115, and 117
(112 and 114)  up to  5,   3, and 1  (1 and  1) units, respectively, with respect to the $xn$ evaporation channels.
In addition, in the nuclei produced the electron capture can
occur by adding one more neutron in the daughter nuclei.
The proton evaporation channels seem to be more
effective to approach  $N=184$ than the alpha emission channels.
One can produce more neutron-rich isotopes
in the reactions with even$-Z$ targets than in the reactions with odd$-Z$ ones.
The   $pxn-$ and $\alpha xn-$evaporation channels allows us to obtain an access to
those isotopes which are unreachable in the $xn-$evaporation channels due to the lack of proper
projectile-target combination.
Thus, employing reactions suggested, one can produce the heaviest isotopes closer
to the center of the island of stability. The  $pxn-$  and  $\alpha xn-$evaporation  channels can be only distinguished by
different $\alpha-$decay chains of the evaporation residues because the excitation functions of these channels overlap
with those from $xn-$evaporation  channels.

Our present results were compared with those obtained with the same fusion model and other mass table
 and with completely other fusion model~\cite{SWCK,SWCK2,SWCK3}
 for which nuclear properties where calculated using the same macroscopic-microscopic
 model but with quasiparticle method for pairing.
Absolute values of cross sections are rather close. However, the ratio of the cross sections in the
maxima of excitation functions is sensitive to the mass table. For example, $\sigma_{p2n}>\sigma_{p3n}$ with the mass table
\cite{moller}, while $\sigma_{p2n}\approx\sigma_{p3n}$ in the calculations presented. If the same mass table is used with
different fusion model, the fusion probability creates the difference in the cross sections obtained.
For example, the ratios $\sigma_{5n}/\sigma_{4n}$ and $\sigma_{\alpha 2n}/\sigma_{\alpha 3n}$ are sensitive to the
increase rate of $P_{CN}$ with excitation energy and, thus, to the fusion model.

\acknowledgments

G.G.A. and N.V.A.  acknowledge  the partial
supports from the Alexander von Humboldt-Stiftung (Bonn).
This work was partly supported by RFBR (17-52-12015, 20-02-00176),    DFG (Le439/6-1), and
Tomsk Polytechnic University Competitiveness Enhancement Program grant.
M.K. was co-financed by the National Science Centre under
Contract No. UMO-2013/08/M/ST2/00257 (LEA-COPIGAL).

\end{document}